\documentclass[fleqn,usenatbib]{mnras}

\usepackage{newtxtext,newtxmath}

\usepackage[T1]{fontenc}
\usepackage{ae,aecompl}

\usepackage{physics}
\usepackage{cleveref}

\usepackage{graphicx}	
\usepackage{amsmath}	
\usepackage{amssymb}	
\usepackage[normalem]{ulem}

\graphicspath{{./figures/}}

\title[Core-Powered Mass-Loss: Stellar Dependence]{Signatures of the Core-Powered Mass-Loss Mechanism in the Exoplanet Population: Dependence on Stellar Properties and Observational Predictions}

\author[A. Gupta and H.E. Schlichting]{
Akash Gupta$^{1}$\thanks{E-mail: akashgpt@ucla.edu} and
Hilke E. Schlichting$^{1,\;2,\;3}$
\\
$^{1}$Department of Earth, Planetary, and Space Sciences, University of California, Los Angeles, CA 90095, USA\\
$^{2}$Department of Physics and Astronomy, University of California, Los Angeles, CA 90095, USA\\
$^{3}$Department of Earth, Atmospheric and Planetary Sciences, Massachusetts Institute of Technology, MA 02139, USA
}

\date{Accepted XXX. Received YYY; in original form ZZZ}

\pubyear{2019}

\begin{document}
\label{firstpage}
\pagerange{\pageref{firstpage}--\pageref{lastpage}}
\maketitle

\begin{abstract}
Recent studies have shown that atmospheric mass-loss powered by the cooling luminosity of a planet's core can explain the observed radius valley separating super-Earths and sub-Neptunes, even without photoevaporation. In this work, we investigate the dependence of this core-powered mass-loss mechanism on stellar mass ($M_\ast$), metallicity ($Z_\ast$) and age ($\tau_\ast$). Without making any changes to the underlying planet population, we find that the core-powered mass-loss model yields a shift in the radius valley to larger planet sizes around more massive stars with a slope given by $\text{d log} R_p/\text{d log} M_\ast \simeq 0.35$, in agreement with observations. To first order, this slope is driven by the dependence of core-powered mass-loss on the bolometric luminosity of the host star and is given by $\text{d log} R_p/\text{d log} M_\ast \simeq (3\alpha-2)/36 \simeq 0.33$, where $(L_\ast/L_\odot) = (M_\ast/M_\odot)^\alpha$ is the stellar mass-luminosity relation and $\alpha\simeq 4.6$ for the CKS dataset. We therefore find, in contrast to photoevaporation models, no evidence for a linear correlation between planet and stellar mass, but can't rule it out either. In addition, we show that the location of the radius valley is, to first order, independent of stellar age and metallicity. Since core-powered mass-loss proceeds over Gyr timescales, the abundance of super-Earths relative to sub-Neptunes increases with age but decreases with stellar metallicity. Finally, due the dependence of the envelope's cooling timescale on metallicity, we find that the radii of sub-Neptunes increase with metallicity and decrease with age with slopes given by $\text{d log} R_p/\text{d log} Z_\ast \simeq 0.1$ and $\text{d log} R_p/\text{d log} \tau_\ast \simeq -0.1$, respectively. We conclude with a series of observational tests that can differentiate between core-powered mass-loss and photoevaporation models.

\end{abstract}


\section{Introduction} \label{sec:intro}

In the last decade, NASA's \textit{Kepler} mission has revolutionized the field of exoplanets by discovering more than 4000 planetary candidates \citep[e.g.,][]{BK10}. These discoveries offer new insights into the formation and evolution of planets. One of $Kepler$'s key findings is that the most common planets, observed to date, are one to four Earth radii ($R_\oplus$) in size \citep[e.g.,][]{FT13,PM13}. In addition, transit-timing variations \citep[e.g.,][]{carter2012a,hadden2017a} and follow-up radial velocity \citep[e.g.,][]{marcy2014a,weiss2014a} measurements have revealed that planets smaller than about 1.6 $R_\oplus$ have higher densities suggesting rocky `Earth-like' compositions while larger planets have lower densities consistent with significant H/He envelopes \citep[e.g.,][]{marcy2014b,rogers2015a}. Intriguingly, analyses of early photometry data from the \textit{Kepler} Input Catalog \citep[e.g.,][]{brown2011a}, the recent spectroscopic follow-up of planet-hosting stars by the  California-\textit{Kepler} Survey \citep[CKS;][]{petigura2017a,johnson2017a}, and the latest studies incorporating \textit{Gaia} astrometry and asteroseismology-based data have all revealed a `{radius valley}' in the distribution of these small, close-in (orbital period $<$ 100 days) exoplanets  \citep[e.g.,][Fulton et al. 2017; Fulton \& Petigura 2018; Van Eylen et al. 2018; Martinez et al. 2019]{owen2013a,LF13,berger2018a}. 

This radius valley marks the transition from a population of small, rocky `super-Earths' to a population of large `sub-Neptunes' with significant H/He envelopes \citep[e.g.,][]{owen2013a,LF13,LF14,rogers2015a,ginzburg2016a}. Furthermore, studies involving \textit{Gaia} astrometry  \citep{martinez2019a} and asteroseismology-based \citep{eylen2018a} high-precision measurements of stellar parameters have measured the slope of the radius valley and obtained values of $\text{d log} R_p/\text{d log} P = -0.11_{-0.03}^{+0.03}$ and $-0.09_{-0.02}^{+0.04}$, respectively.

The bimodality in the radius distribution has been attributed to {photoevaporation} of H/He envelopes by high-energy radiation (e.g., XUV) from the host stars \citep[e.g.,][]{owen2013a,LF13,jin2014a,chen2016a,owen2017a,jin2018a}. Under this mechanism, close-in planets can receive, over a lifetime, a total time-integrated high-energy flux comparable to their atmosphere's binding energy and are thus able to lose their atmospheres \citep[e.g.,][]{owen2017a}. Studies have demonstrated that planetary evolution with photoevaporation can reproduce the radius valley. In addition, these studies infer `Earth-like' bulk composition of the planetary cores implying a low overall water-fraction \citep[e.g.,][]{owen2017a,jin2018a,wu2019a}.

Recently, \citet{ginzburg2018a} and \citet{gupta2019a} showed that the {core-powered mass-loss} mechanism is also able to reproduce the observed radius valley, even in the absence of photoevaporation. In this mechanism, a planet's internal luminosity drives the loss of its atmosphere \citep{ginzburg2016a,ginzburg2018a,gupta2019a}. The source of this luminosity is a planet's primordial energy from formation, which can be of the order of, or even larger than, its atmosphere's gravitational binding energy. \citet{gupta2019a} showed that evolution of planets under the core-powered mass-loss mechanism around `Sun-like' host stars {, i.e. stars with solar mass and metallicity and an age of 3 Gyrs,} successfully reproduces the location, shape and slope of the radius valley as a function of orbital period. They also demonstrated that the exact location of the radius valley depends on the bulk composition of the planetary cores whereas the relative abundance of super-Earths and sub-Neptunes depends on the underlying planet mass distribution. Similar to photoevaporation studies, \citet{gupta2019a} found that the observations are consistent with `Earth-like' cores with a maximum water-fraction of $\sim$20\%.

New observations provide us with the opportunity to characterize the super-Earth and sub-Neptune populations as a function of their host star properties \citep[e.g.,][]{fulton2018a,petigura2018a,dong2018a}. This opens new avenues to investigate and contrast possible signatures the core-powered mass-loss mechanism and photoevaporation imprint on the exoplanet population \citep[e.g.,][]{owen2018a,wu2019a}.

\citet{fulton2018a} and \citet{wu2019a} have found that the radius valley increases in planet size with stellar mass. Moreover, \citet{fulton2018a} found that the planet size distribution also shifts to higher insolation flux with increasing stellar mass.

In addition, \citet{petigura2018a} have reported that the relative occurrence of sub-Neptunes increases with increasing stellar metallicity while the relative occurrence of super-Earths remains largely unchanged; see also \citet{dong2018a}. Furthermore, \citet{dong2018a} and \citet{owen2018a} have reported that sub-Neptunes are larger around higher metallicity FGK stars; see also \citet{petigura2018a}. \citet{hirano2018a} reported a similar trend in the distribution of planet sizes around M-dwarfs.

Furthermore, studies based on the Zodiacal Exoplanets in Time (ZEIT) survey have reported a trend in planet size with stellar age \citep[e.g.,][]{mann2016a,rizzuto2018a}. Most \textit{Kepler} planets are older than a Gyr and have significant uncertainties in their age estimates. This makes it difficult to investigate any trends in planet size with stellar age. However, observations from the \textit{K2} mission have allowed the discovery of planets in young clusters like Praesepe and Hyades, both of which are about 650 million years old \citep[e.g.,][]{mann2016a,rizzuto2018a}. The number of planets observed in such clusters is still below ten and thus not statistically significant, nevertheless, these younger planets are bigger, on average, than their older \textit{Kepler} counterparts; see Figure 12 in \citet{rizzuto2018a}.

In this study, we extend previous work on the core-powered mass-loss mechanism \citep{ginzburg2018a,gupta2019a}, and investigate how stellar mass, metallicity and age impact the resulting planet size distribution. We model our host star population on the stars in the CKS survey and investigate, collectively and in isolation, the impact of stellar mass, metallicity and age on the resulting exoplanet size distribution. We show that core-powered mass-loss can explain many of the observational trends with stellar properties discussed above. In addition, we quantify how different stellar properties shape the observed planet size distribution. {In contrast} to photoevaporation models, we find no evidence for a  {linear} correlation between planet and stellar mass {but can't rule it out either.} We show that {in the core-powered mass-loss model} the observed dependence of the radius valley on stellar mass is, to first order, driven by the stellar mass-luminosity relation.

This paper is structured as follows: In Section \ref{sec:model}, we describe the core-powered mass-loss mechanism and discuss our model for planetary structure and composition. We then explain, in Section \ref{sec:model_population}, how we model the exoplanet and stellar populations. The numerical and analytical core-powered mass-loss results are divided into two sections. {In Section \ref{sec:results_comparison_w_obs}, we present a comparison of our results with the current exoplanet observations. Complementing this, in Section \ref{sec:results_stellar_prop}, we show how the core-powered mass-loss mechanism depends on individual stellar parameters, make predictions for trends with stellar mass, metallicity and age and investigate if there is any correlation between planet and stellar mass.} Finally, we summarize our results in Section \ref{sec:conclusion} and present observational tests for distinguishing between the signatures of core-powered mass-loss and photoevaporation.

\section{Planet Structure and Evolution under the Core-Powered Mass-Loss Mechanism}\label{sec:model}

To study the imprint of the core-powered mass-loss mechanism on the exoplanet population, we solely focus on core-powered mass neglecting mass-loss due to photoevaporation and other mechanisms. For a detailed review of this mechanism, we refer the reader to \citet{ginzburg2016a} and for its dependence on planetary properties, to \citet{gupta2019a}.

As a planetary core grows by accreting solids, gravitational binding energy is converted into heat. This thermal energy can be efficiently radiated away if the core forms in isolation. However, if this accretion occurs in the presence of a gas disk, then the core will start accreting a H/He envelope from the surrounding nebula once its Bondi radius is larger than its physical radius. The presence of an optically thick envelope acts as a `thermal blanket' since, from this point onward, the core's heat loss is limited by the thermal diffusion across the radiative-convective boundary of the envelope \citep[e.g.,][]{rafikov2006a,lee2015a,ginzburg2016a}, significantly reducing the cooling rate of the underlying core. As a result, the core temperature is essentially set by the maximum temperature that permits the accretion of a H/He envelope and is roughly given by $T_c \sim G M_c \mu/k_B R_c$, where $\mu$ is the mean molecular mass of the atmosphere, $k_B$ is the Boltzmann constant, $G$ is the gravitational constant and $M_c$ and $R_c$ are the mass and radius of the planetary core, respectively. This implies typical core temperatures of 10$^4$ - 10$^5$ K for core masses ranging from the mass of Earth to Neptune. It is this primordial energy from planet formation that drives atmospheric loss in the core-powered mass-loss mechanism which proceeds over Gyr timescales.

As the protoplanetary disk disperses, it causes a loss of pressure support on the outside of the planet's atmosphere. As a consequence, the planet experiences `spontaneous' atmospheric mass-loss driven by the luminosity of the inner regions of its atmosphere \citep{ginzburg2016a,owen2016a}. Due to this atmospheric mass-loss, the atmosphere rapidly shrinks to a few times the planet's core radius over the disk dispersal timescale. This sets the initial condition for planetary evolution under the core-powered mass-loss and photoevaporation \citep{owen2017a} mechanisms.

At this stage, we assume that a typical planet with radius $R_p$ and mass $M_p$ has a dense core of radius $R_c$ and mass $M_c$, and a surrounding gaseous atmosphere of mass $M_{atm}$. We define the core as the non-gaseous part of the planet which dominates the planet's mass such that $M_c \sim M_p$. We account for gravitational compression by assuming that the mass-radius relation of the core is given by ${M_c}/{M_\oplus} = (R_c/R_\oplus)^4(\rho_{c*}/\rho_\oplus)^{4/3}$, where $\rho_{c*}$ is the {density of the core scaled to an Earth mass} and $\oplus$ refers to the corresponding Earth values \citep{valencia2006a,fortney2007a}. Motivated by direct observational measurements of super-Earth masses and radii \citep[see Figure 10 from][]{bower2019} and core-powered mass-loss results from \citet{gupta2019a}, we assume throughout this paper a density of $\rho_{c*}=5.0$ g cm$^{-3}$ for the cores, when scaled to an Earth mass{; see also \citet{dorn2019a,dressing2015a}}. This density estimate is 10\% lower than that used by \citet{gupta2019a} but is in better agreement with the radius distribution observations from \citet{fulton2017a} when considering a distribution of planet host stars modelled after the CKS dataset \citep[][see Section \ref{sec:model_star} for more details]{johnson2017a,fulton2017a}.  {In addition, we assume that the core is molten and fully convective, and that the core-envelope interface is well-coupled.}

Following previous work on gas accretion and loss during disk dispersal \citep[][]{ginzburg2016a}, we model all planets with an initial atmosphere to core mass-fraction ($f$) given by
\begin{equation}\label{eq:f}
f \simeq 0.05 (M_c/M_\oplus)^{1/2}.
\end{equation}
We assume that all planets have initial $H_2$ envelopes and that they can be described as a diatomic ideal gas with $\gamma=7/5$ and $\mu=2 \text{ amu}$. Although the initial atmosphere to core mass-fraction is only a few percent, the presence of the atmosphere significantly increases the size of the planet. Following past works \citep[e.g.,][]{PY14,lee2015a,IS15}, we assume that the atmosphere is structured such that it has an inner convective and outer isothermal region. The convective region is modeled to be adiabatic and contains most of the atmospheric mass while the radiative region is close to isothermal. The atmosphere transitions from the convective to radiative region at the radiative-convective boundary, $R_{rcb}$. We assume that the planet size is given by the radiative-convective boundary, i.e., $R_p \sim R_{rcb}$, since the atmospheric density decreases exponentially outside the $R_{rcb}$. In other words, we assume that the atmospheric thickness is $\Delta R \sim R_{rcb}-R_c$, where $\Delta R$ initially is a few $R_c$.

The aforementioned model yields an atmospheric mass of
\begin{equation}\label{eq:M_atm}
M_{atm} = \frac{\gamma -1}{\gamma} 4\pi R_c^2 \rho_{rcb} {\Delta R} \left( \frac{R_B^\prime {\Delta R} }{R_c^2}\right)^{1/(\gamma -1)}.
\end{equation}
Here $\gamma$ is the ratio of heat capacities for the atmosphere, $\rho_{rcb}$ is the density of the atmosphere at the $R_{rcb}$ and $R_B^\prime$ is the modified Bondi radius, such that
\begin{equation}
R_B^{\prime} \equiv \frac{\gamma - 1}{\gamma} \frac{G M_c \mu}{k_B T_{rcb}},
\end{equation}
where $T_{rcb}$ is the temperature at the $R_{rcb}$. $T_{rcb}$ is approximately the planetary equilibrium temperature, $T_{\text{eq}}$, for a given distance from the host star which can written as
\begin{equation}
    T_{\text{eq}} = \left(\frac{1}{16 \pi \sigma}\frac{L_\ast}{ a^2}\right)^{1/4},
\end{equation}
where $\sigma$ is the Stefan-Boltzmann constant, $a$ is the planet's semi-major axis and $L_\ast$ is the luminosity of the host star. To facilitate direct comparison with observations, we present some of our results as a function of stellar insolation flux, $S$, which is related to the equilibrium temperature by $S/S_\oplus = (T_{\text{eq}}/T_{eq, \earth})^4$.

The total energy available for cooling ($E_{cool}$) is the sum of the thermal energy in the atmosphere and core and the atmosphere's gravitational energy, i.e.
\begin{equation}\label{eq:E_cool}
E_{cool} \simeq g \Delta R \left( \frac{\gamma}{2\gamma -1}M_{atm} + \frac{1}{\gamma}\frac{\gamma-1}{\gamma_c -1} \frac{\mu}{\mu_c} M_c \right),
\end{equation}
where $\mu_c$ and $\gamma_c$ are the molecular mass and adiabatic index of the core, and $g=GM_c/R_c^2$. Since the core-envelope interface is well coupled, the cooling of both the core and atmosphere is dictated by the radiative diffusion through the radiative-convective boundary, and thus the luminosity of the planet is
\begin{equation}\label{eq:L}
L = -\frac{\text{d}E_{cool}}{\text{d}t}= \frac{64 \pi}{3} \frac{\sigma T_{rcb}^4 R_B^{\prime}}{\kappa \rho_{rcb}},
\end{equation}
where $\kappa$ is the opacity at the radiative-convective boundary. We assume that this opacity scales as 
\begin{equation}\label{eq:opacity}
    \frac{\kappa}{0.1\;\text{cm}^2\;\text{g}^{-1}} = \frac{Z_\ast}{Z_{\sun}} \left( \frac{\rho_{rcb}}{10^{-3} \;\text{g}\;\text{cm}^{-3}} \right)^{\beta}
\end{equation}
where $Z_\ast$ is the stellar metallicity, $Z_{\sun}$ the metallicity of the Sun, $\beta$ is assumed to be 0.6 throughout this paper \citep[e.g.,][]{lee2015a,freedman2008a}  {and any temperature dependence is neglected}. 
We are assuming in Equation \ref{eq:opacity} that the envelope opacity scales linearly with the metallicity of the envelope and that the envelope metallicity can be approximated by the metallicity of the host star such that $[\text{Fe/H}] \simeq [Z_\ast/Z_{\sun}]$.  {The assumption that the envelope metallicity scales linearly with the metallicity of the host star is a reasonable first guess, given the absence of direct observational measurements, since more metal-rich stars are expected to have more metal-rich disks which should, in turn, lead to more metal-rich planetary envelopes. Recent observational measurements provide tentative evidence for a correlation between the 
metallicity of envelope and that of the host star for gas giant exoplanets \citep[e.g.,][]{wallack2019a}.}

Combining \Cref{eq:E_cool,eq:L} yields a planetary cooling timescale given by
\begin{equation}\label{eq:t_cool}
t_{cool}= \frac{E_{cool}}{|\text{d}E_{cool}/\text{d}t|}=\frac{E_{cool}}{L}.
\end{equation}

\begin{figure*}
\centering
\includegraphics[width=\textwidth,trim=310 580 290 350,clip]{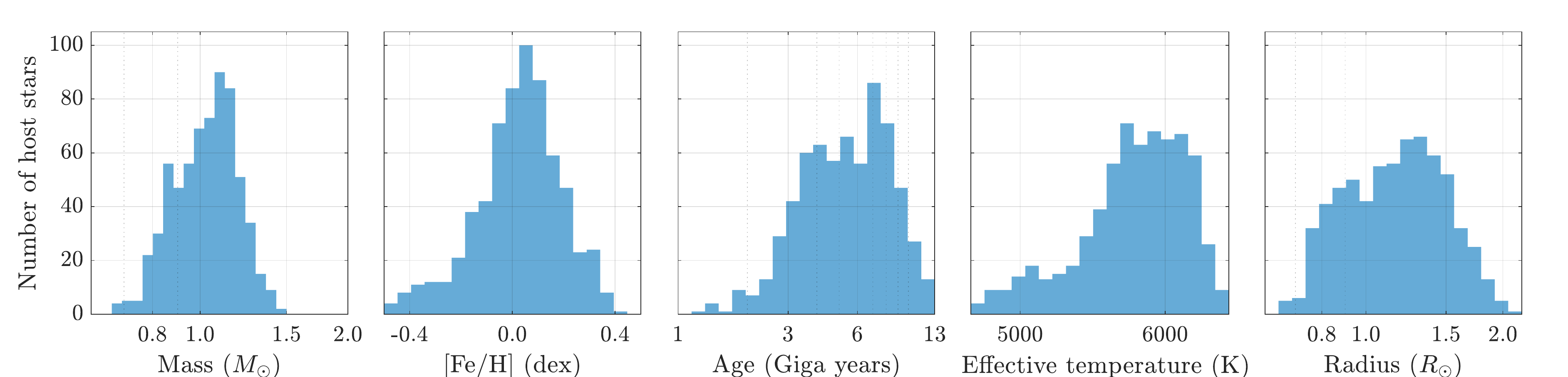} 
\caption{Distributions of stellar properties from the CKS dataset  {\citep[][see discussion in Section \ref{sec:model_star}]{petigura2017a,johnson2017a,fulton2017a}}.}
\label{fig:CKS_data}
\end{figure*}

There are two regimes of atmospheric mass-loss: energy-limited mass-loss and Bondi-limited mass-loss. The energy-limited mass-loss rate is
\begin{equation}\label{eq:M_loss_rate_E}
\dot{M}_{atm}^E \simeq \frac{L(t)}{g R_c},
\end{equation}
i.e., it is the absolute upper limit on the atmospheric mass-loss rate achievable assuming all the cooling luminosity goes into driving the mass-loss. On the other hand, the Bondi-limited mass-loss rate, $\dot{M}_{atm}^B$, signifies a physical limit on the atmospheric mass-loss rate dictated by the thermal velocity of the gas molecules at the Bondi radius \citep[e.g.,][]{ginzburg2016a,owen2016a}. {Because the hydrodynamic outflow of the atmosphere has to pass through the sonic point and since the mass flux is conserved, it is convenient to determine the mass-loss rate at the sonic point, $R_s=GM_p/2c_s^2$, where $c_s = (k_B T_{\text{eq}}/\mu)^{1/2}$ is the isothermal speed of sound. In this case, the mass-loss rate is $\dot{M}=4 \pi \rho_s R_s^2 c_s$, where $\rho_s$ is the density at the sonic point.} In the limit that $R_s>>R_{rcb}$, $\rho_s$ is related to the density at the radiative-convective boundary by $\rho_s = \rho_{rcb} \text{exp} (-2R_s/R_{rcb})$. The mass-loss rate at the Bondi radius is therefore given by 
\begin{equation}\label{eq:M_loss_rate_B}
\dot{M}_{atm}^B = 4\pi R_s^2 c_s \rho_{rcb} \; \text{exp}\left( -\frac{GM
_p}{c_s^2 R_{rcb}}\right).
\end{equation}
The actual atmospheric mass-loss rate is the minimum of the energy-limited and Bondi-limited mass-loss rates, i.e., $\dot{M}_{atm}=min\{\dot{M}_{atm}^E,\dot{M}_{atm}^B\}$. Combining \Cref{eq:M_atm,eq:M_loss_rate_E,eq:M_loss_rate_B} yields an atmospheric mass-loss timescale that can be written as
\begin{equation}\label{eq:t_loss}
t_{loss}=\frac{M_{atm}}{|\text{d}M_{atm}/\text{d}t|} = max\left\{ \frac{M_{atm}}{\dot{M}_{atm}^E}, \frac{M_{atm}}{\dot{M}_{atm}^B} \right\}.
\end{equation}

Planets which eventually become super-Earths have enough energy to lose their atmospheres entirely and have mass-loss timescales shorter than cooling timescales. On the other hand, sub-Neptunes either do not have enough energy for complete atmospheric loss or have mass-loss timescales that are longer than their cooling timescales or the age of the system.

In our numerical simulations, we simultaneously calculate cooling and atmospheric mass-loss of a planet by evolving the planet's energy and atmospheric mass as
\begin{align}
E_{cool}(t+\text{d}t) &=  E_{cool}(t) - L(t) \text{ d}t \text{ and}\\
M_{atm}(t+\text{d}t) &= M_{atm}(t) - \dot{M}_{atm}(t) \text{ d}t,
\end{align}
respectively; using \Cref{eq:M_atm,eq:E_cool,eq:M_loss_rate_E,eq:M_loss_rate_B}. We choose an integration time step dt = $10^{-2}$ $\times$ min$\left\lbrace t_{cool}, t_{loss} \right\rbrace $.

\section{Modeling the exoplanet population and its host stars} \label{sec:model_population}
In this section, we discuss how we model the planetary and stellar populations in our numerical simulations. 

\subsection{Exoplanet Population}\label{sec:model_planet}
Following  {\citet{owen2017a} and \citet{gupta2019a}}, we adopt the following orbital period and planet mass distribution to model the exoplanet population:
\begin{equation}{\label{eq:P_distr}}
\dv{N}{\;\text{log}P} \propto \begin{cases}
     P^{2}, & {P < 8\; \text{days}} \\
    \text{constant}, & {P > 8\; \text{days}}, \text{ and} 
  \end{cases}  
\end{equation}
\begin{equation}{\label{eq:M_c_distr}}
\dv{N}{M_c} \propto \begin{cases}
    M_c\;\text{exp} \left( -{M_c^2}/{(2 \sigma_{M}^2)} \right),&{M_c < 5\; M_\oplus}  \\
    M_c^{-2},&{M_c > 5\; M_\oplus}.
  \end{cases}  
\end{equation}
Here, $\sigma_M$ is the planet mass where the planet mass distribution peaks. As in \citet{gupta2019a}, we assume $\sigma_M=3 M_\oplus$ unless otherwise stated. Throughout this work, we assume that the period distribution is independent of stellar mass, which is, to first order, consistent with observations \citep{fulton2018a}.

\subsection{Host Star Population}\label{sec:model_star}

We model the stellar population after the full CKS dataset  {\citep{petigura2017a} and use the isochrone-fitted stellar parameters from \citet{johnson2017a}}.  {The full CKS dataset consists of 2025 \textit{Kepler} planets. We exactly follow \citet{fulton2017a} to implement a series of cuts to restrict this dataset to a well-characterized sub-sample of 900 planets and 652 host stars. These cuts involve removing false-positive planetary candidates, limiting the sample to planets with impact parameter $b$ $<$ 0.7 and orbital periods $P$ $<$ 100 days, and excluding faint stars with Kepler-magnitudes $K_p$ $>$ 14.2, giant stars \citep[using the empirical criteria mentioned in][]{fulton2017a} and stars with effective temperatures $T_{\text{eff}}$ $<$ 4700 K and $T_{\text{eff}}$ $>$ 6500 K. Hereafter, we simply refer to this refined subset of host stars as the CKS dataset or CKS stellar distribution.}

We characterize a host star by the following properties: mass ($M_\ast$),  {radius ($R_\ast$), effective temperature ($T_{\text{eff}}$)}, metallicity ($Z_\ast$) and age ($\tau_\ast$). \Cref{fig:CKS_data} shows histograms of these stellar properties from the CKS dataset. We use these parameters as stellar inputs when numerically modelling a planet's thermal evolution and mass-loss while further assuming that the age of the planets is well approximated by the age of their host stars. In total, we evolve a population of 10,000 planets with varying periods and masses (see \Cref{eq:P_distr,eq:M_c_distr}) around each of the  {652} CKS stars.  {The stellar parameters are kept fixed in time, as the stars in our sample are not expected to evolve significantly over the typical core-powered mass-loss timescales, which range from 0.5 to a few Gyrs. We present the results from our core-powered mass-loss model for a host star population modeled after the CKS dataset and compare them with observations in \Cref{sec:results_comparison_w_obs}.}

In addition, in order to understand the impact of the individual stellar parameters on the observable exoplanet population, we also investigate how the core-powered mass-loss depends on each of the stellar parameters (mass, metallicity and age) separately in \Cref{sec:results_stellar_prop}. For the stellar parameters that we do not vary, we assume $M_\ast=1\;M_\odot$, $[Z_\ast/Z_\odot]=\;0.0$ dex and $\tau_\ast =$ 3 Gyrs. As in \Cref{sec:results_comparison_w_obs}, we evolve a population of 10,000 planets with period and mass distributions given by \Cref{eq:P_distr,eq:M_c_distr} around each of the host stars in the population.

\begin{figure}
\centering
\includegraphics[width=0.4\textwidth,trim=600 340 670 530,clip]{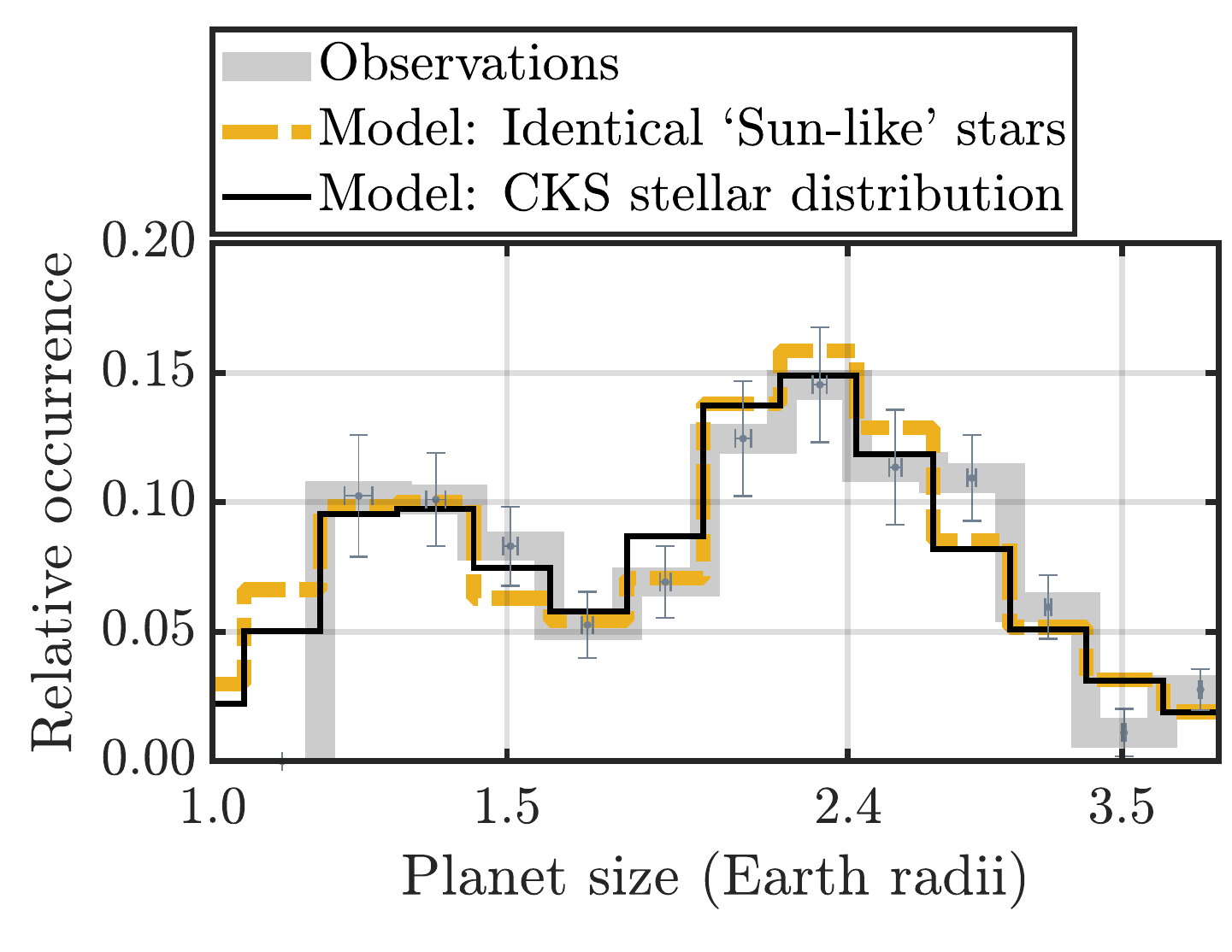} 
\caption{ {Comparison of results based on the core-powered mass-loss evolution model with observations in a one-dimensional histogram of planet size. The thick-gray histogram corresponds to completeness-corrected observations by \citet[][see Table 3]{fulton2017a} for planets larger than 1.16 $R_{\oplus}$. The orange-dashed histogram corresponds to the results from \citet{gupta2019a} assuming planetary evolution around identical `Sun-like' host stars. The thin-black histogram corresponds to results of our model assuming planetary evolution around a distribution of host stars based on the CKS dataset \citep{johnson2017a,fulton2017a}. The results show good agreement with the observations.}}
\label{fig:1D_comparison}
\end{figure}

\section{Results: Comparison with Observations} \label{sec:results_comparison_w_obs}
In this section, we compare our core-powered mass-loss results with observations of the planet size distribution and its dependence on orbital period, {insolation flux and} stellar mass{,} metallicity {and age}.

\citet{ginzburg2018a} and \citet{gupta2019a} demonstrated that, for a population of planets around  {identical} {`Sun-like'} stars, the core-powered mass-loss mechanism successfully reproduces the radius valley observed in the distribution of small, close-in exoplanets. Here, we extend these {previous works} to account for the actual, observed properties of the host star population. All the results presented in this section are based on a host star population modeled after the stellar properties of the CKS dataset; see Section \ref{sec:model_star} for details. This facilitates direct comparison between our results and observations based on the CKS dataset \citep[e.g.,][]{fulton2017a,fulton2018a,owen2018a,wu2019a}.

\begin{figure*}
\centering
\includegraphics[width=.33\textwidth,trim=290 455 895 370,clip]{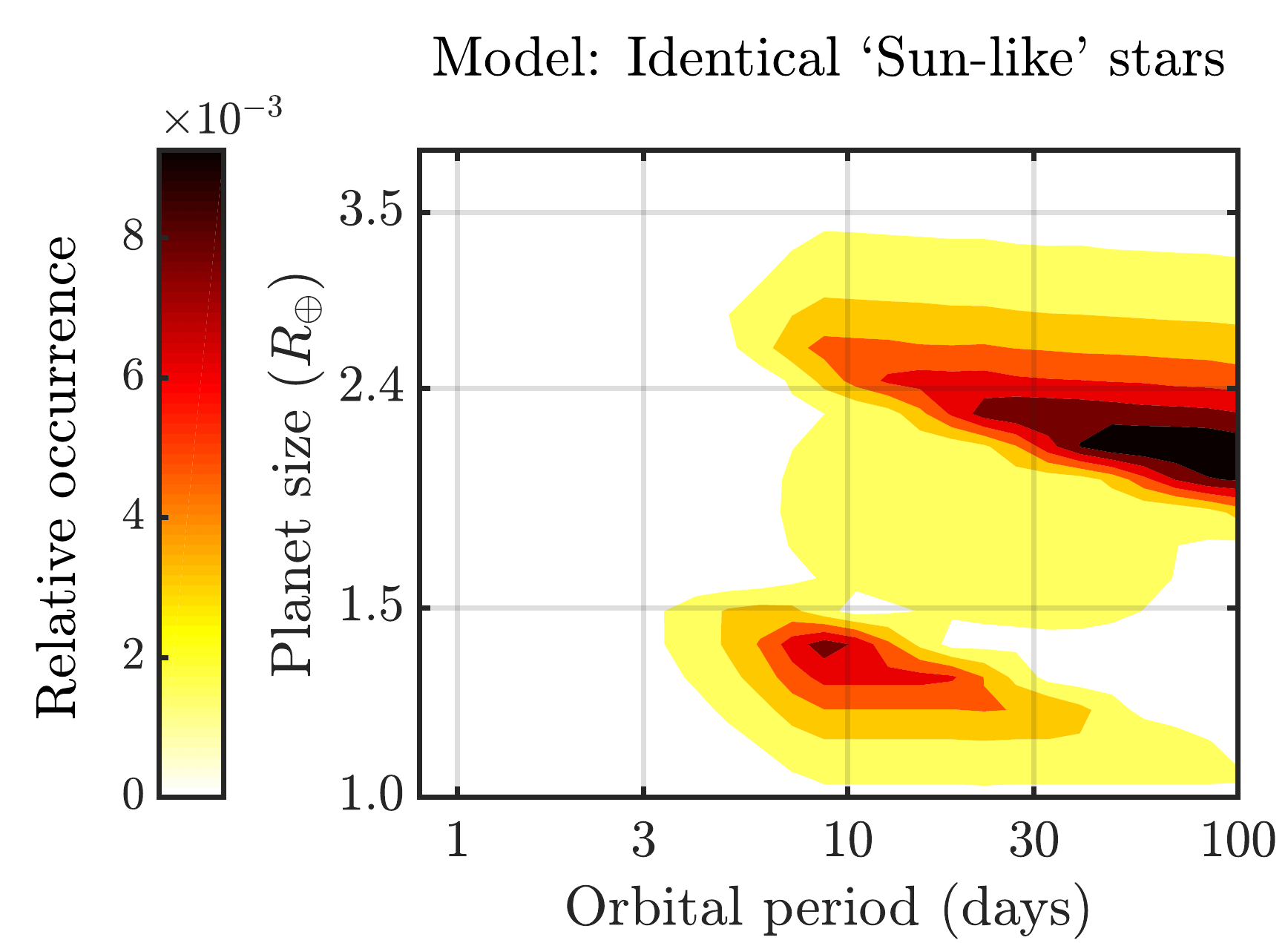}
\includegraphics[width=.33\textwidth,trim=820 455 375 370,clip]{PVsR_p_new.pdf} 
\includegraphics[width=.315\textwidth,trim=0 2 0 0,clip]{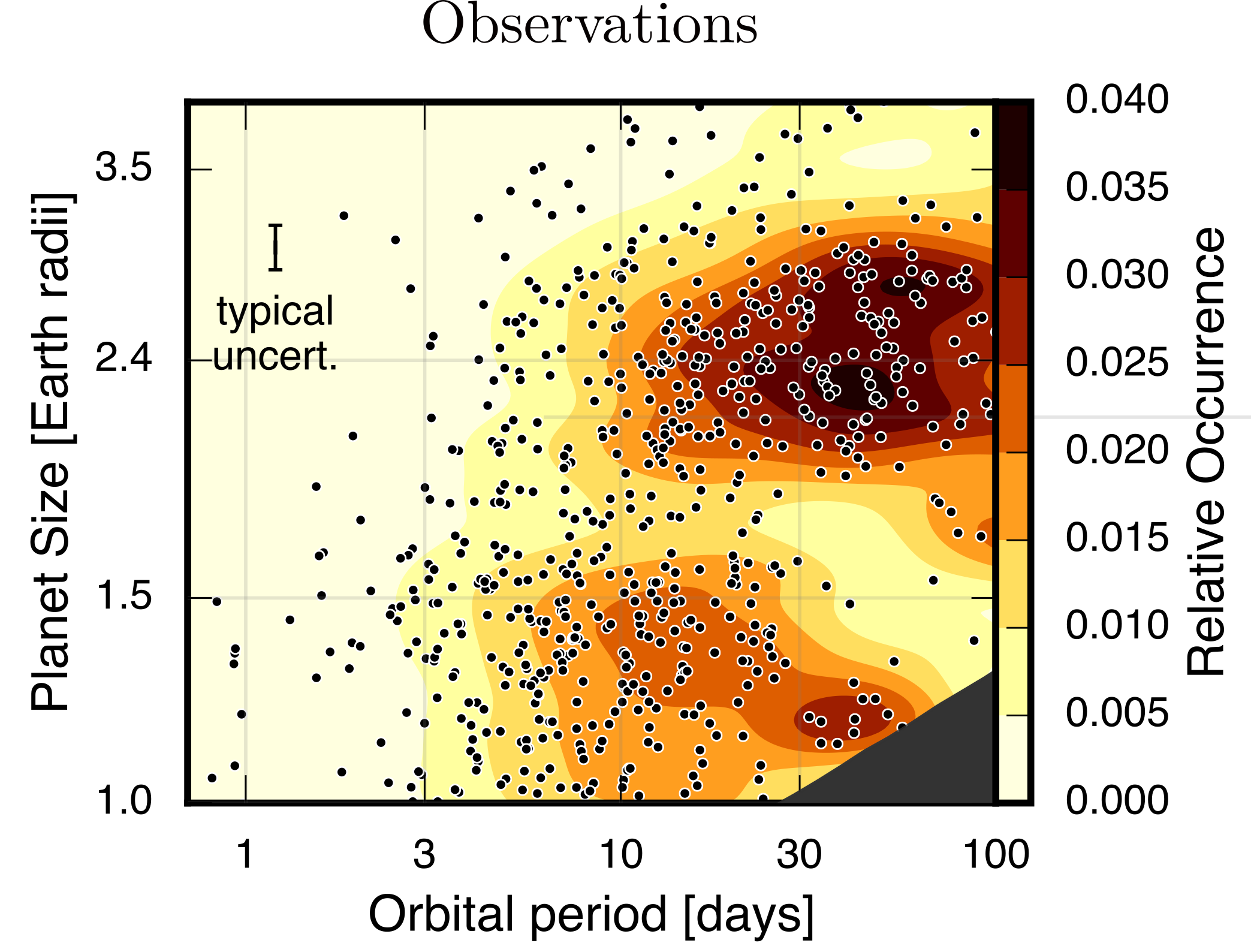}
\includegraphics[width=.33\textwidth,trim=290 455 895 395,clip]{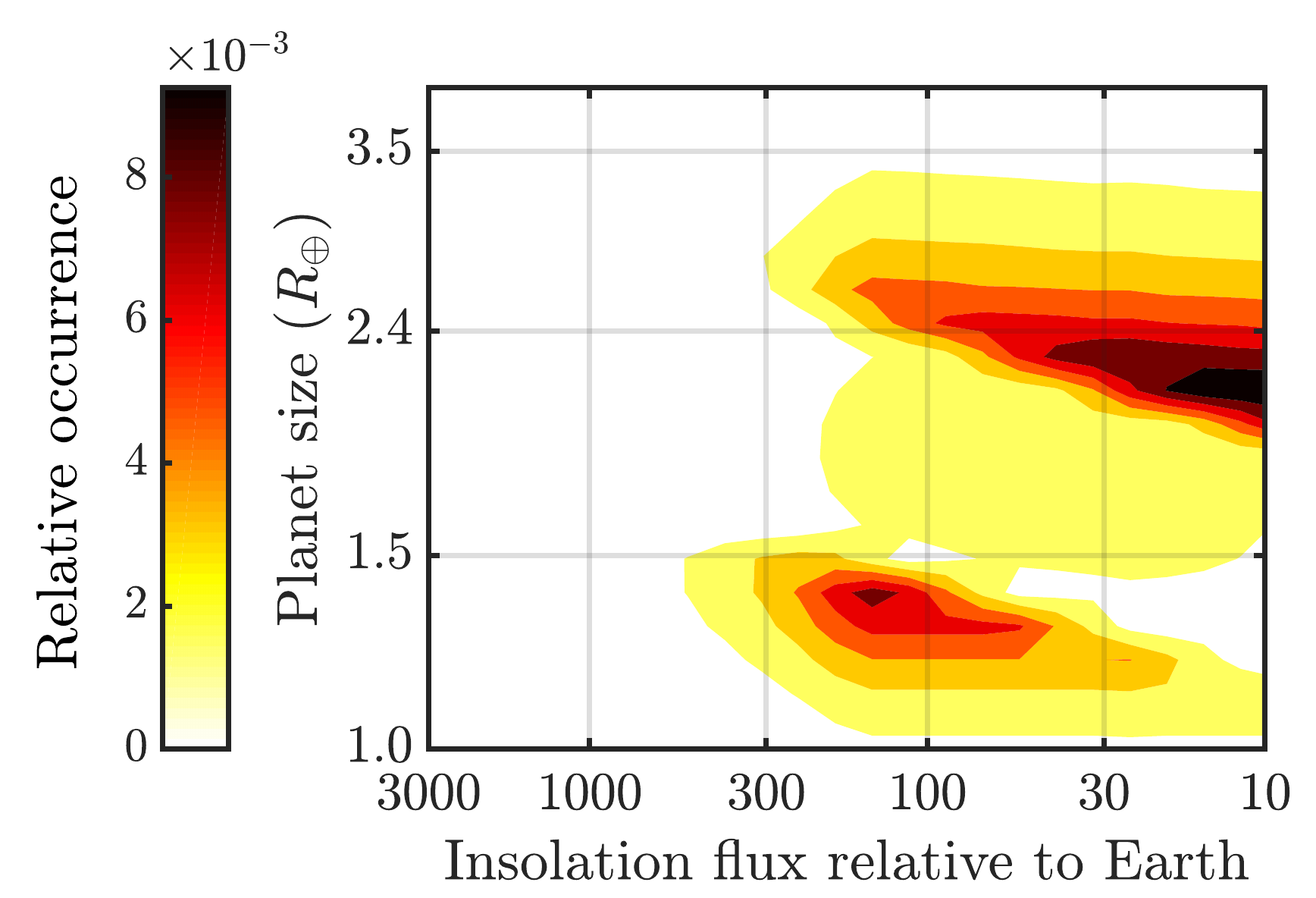}
\includegraphics[width=.33\textwidth,trim=820 455 375 395,clip]{SVsR_p_new.pdf}
\includegraphics[width=.315\textwidth,trim=0 2 0 0,clip]{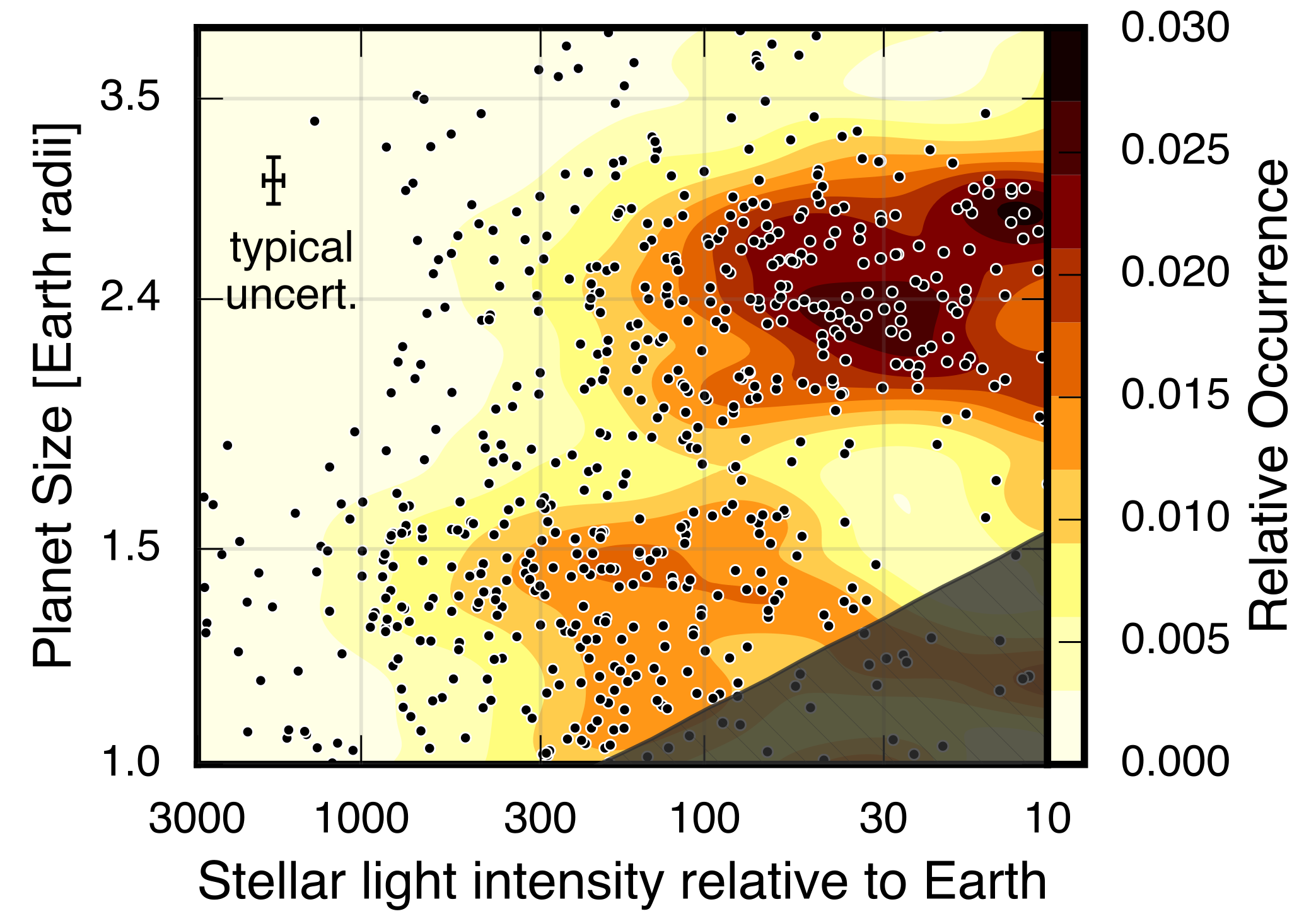}
\caption{Comparison of core-powered mass-loss results with observations.
The top row shows planet size as a function of orbital period whereas the bottom row displays planet size as a function of stellar insolation flux. The left panel corresponds to results from \citet{gupta2019a} which were calculated for planets around identical {`Sun-like'} host stars, the middle panel displays results from this work for planets around host stars that are modeled after the stellar properties of the CKS dataset \citep{johnson2017a,fulton2017a}, and the right panel corresponds to {completeness-corrected} observed planet size distribution from \citet{fulton2018a} based on the CKS dataset (reproduced with permission). 
}
\label{fig:2D_comparison}
\end{figure*}

\begin{figure*}
\centering
\includegraphics[width=.53\textwidth,trim=275 450 880 370,clip]{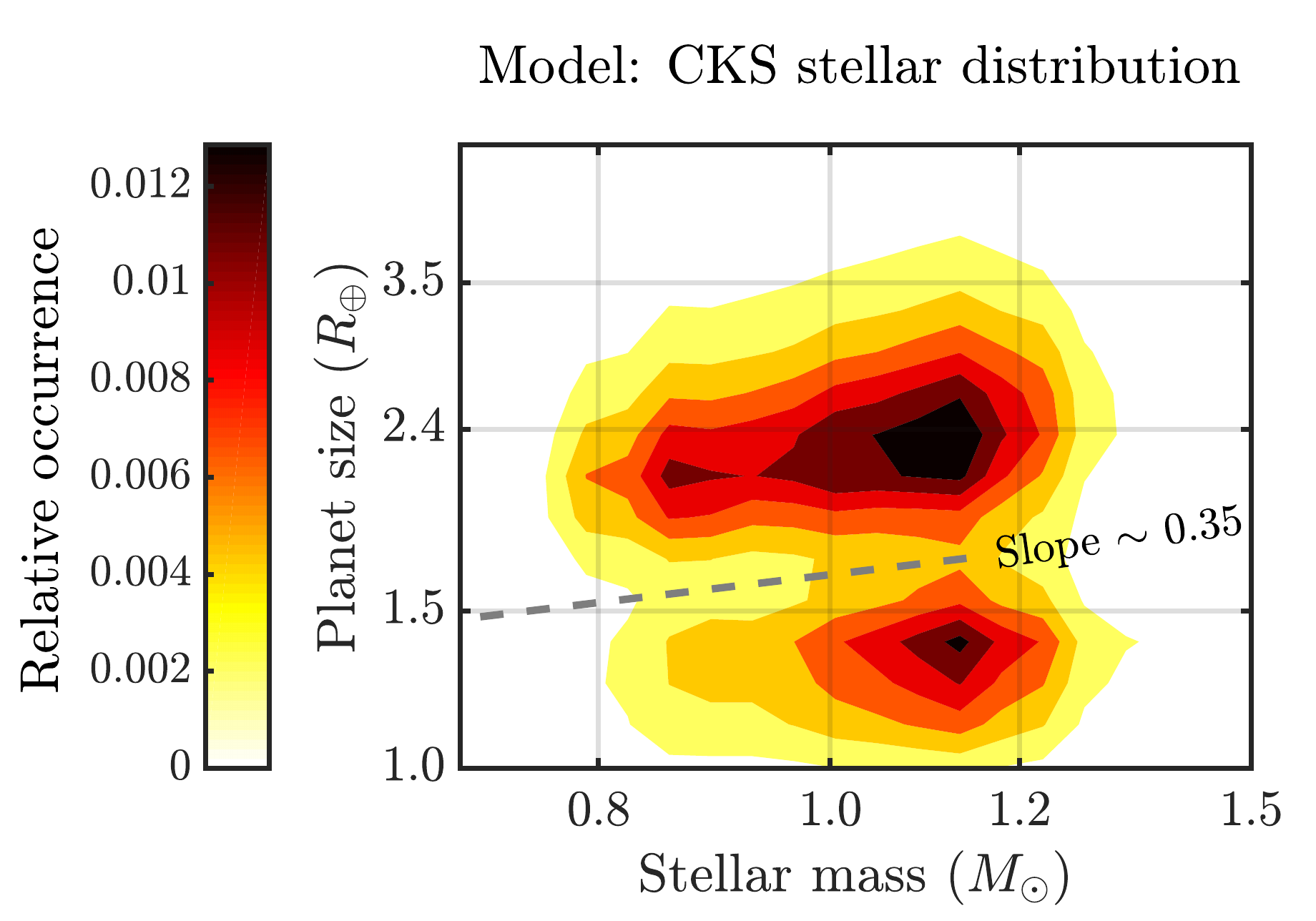}\hfill \includegraphics[width=.46\textwidth,trim=0 0 0 0,clip]{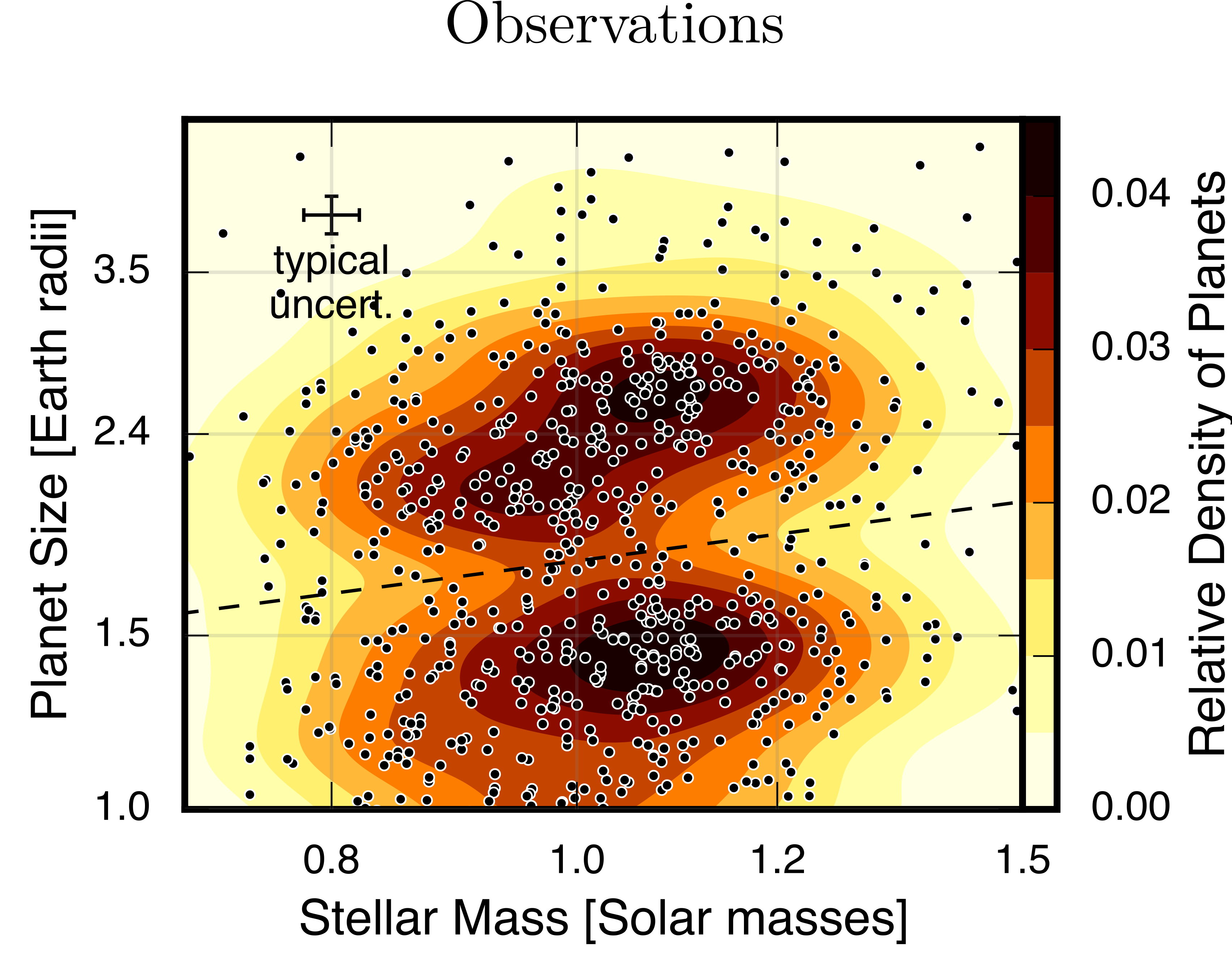} 
\caption{Comparison of core-powered mass-loss results (left-panel) with observations (right-panel) from \citet{fulton2018a} (reproduced with permission). The left panel shows the dependence of the core-power mass loss results on stellar mass, demonstrating that larger planetary cores are stripped of their envelopes when residing around more massive host stars compared to planets around lower mass stars. This yields a shift in the valley separating the super-Earth and sub-Neptune populations to larger planet radii for higher mass stars. To first order, this slope is driven by the dependence of the core-powered mass-loss mechanism on the bolometric luminosity of the host star as this dictates the outer boundary condition for atmospheric mass-loss. The dashed lines in both panels indicate the shift in the valley's location with stellar mass and are plotted with the same slope. {\textit{Note:} Completeness corrections associated with observational bias were not accounted for in the right panel (B.J. Fulton, personal communication). Therefore, the relative occurrence of planets above and below the valley cannot be compared with the observations.}}
\label{fig:comparison_prop_m}
\end{figure*}

{In \Cref{fig:1D_comparison,fig:2D_comparison,fig:comparison_prop_m_bins,fig:comparison_prop_Z}, we compare our results with completeness-corrected observations (right-hand panels) from \citet{fulton2017a}, \citet{fulton2018a} and \citet{owen2018a}. These studies make completeness-corrections to recover the underlying planet size distribution from the distribution of detected planets. To achieve this, they account for transit and detection probabilities when calculating the underlying planet occurrence rates. While the transit probability simply depends on the ratio of host star radius and the planet's semi-major axis, the detection probability depends on the signal-to-noise ratio, planet-to-star size ratio and orbital period and is estimated by injecting synthetic transits into the Kepler data to check if they can be recovered or not \citep{christiansen2015a}; for details, see \citet{fulton2017a}. In contrast, the observations from \citet{fulton2018a} shown in the right-hand panel of \Cref{fig:comparison_prop_m,fig:stellar_prop_var_M_2} have not been completeness-corrected such that the relative intensities of the peaks above and below the radius valley are not representative of the underlying planet population.}

{\Cref{fig:1D_comparison} shows a comparison between our core-powered mass-loss results for a host star population modeled on the CKS dataset and with results obtained for `Sun-like' host stars from \citet{gupta2019a} and completeness-corrected observations reported by the CKS team \citep[e.g.][]{fulton2017a}. The histogram shows no significant difference between results obtained for `Sun-like' host stars (dashed orange-line) and stars modeled directly after the CKS dataset (thin black-line). Both models show good agreement with the observations (thick grey-line).}

\subsection{Changes in planet size distribution with orbital period and insolation flux}

\Cref{fig:2D_comparison} shows a comparison{, in planet size-orbital period and insolation flux space,} of our core-powered mass loss results for CKS host stars (middle-panel), with results based on identical {`Sun-like'} host stars \citep[left-panel;][]{gupta2019a} and observations \citep[right-panel;][]{fulton2018a}. The core-powered mass-loss results are in good agreement with observations successfully reproducing the location, shape and slope of the valley, as well as the location and magnitude of the peaks of the exoplanet populations above and below the valley. In addition, comparison of the middle- and left-panel shows that the core-powered mass-loss results for the two stellar populations are similar but, accounting for the true properties of the underlying stellar population (middle-panel) reproduces even finer features of the observations (right-panel). For example, our new results can reproduce the `triangular' shape of the radius valley seen in the observations (right-panel) better than our previous study that was based on identical {`Sun-like'} host stars \citep{gupta2019a}. This is due to the range of stellar parameters used in this work, which leads to a shallower upper edge of the valley compared to results for planets around identical {`Sun-like'} stars.

\subsection{Changes in planet size distribution with stellar mass}

\Cref{fig:comparison_prop_m} shows the resulting distribution of planet size as a function of stellar mass from our core-powered mass-loss model (left-panel) and observations from \citet[][right-panel]{fulton2018a}. We find that core-powered mass-loss causes the location of the radius valley and the super-Earth and sub-Neptune populations to increase in planet size with increasing stellar mass. This is consistent with observational results from \citet{fulton2018a} and \citet{wu2019a}. In addition, we find that the slope of the valley in this parameter space is $\text{d log} R_p/ \text{d log} M_\ast \sim  0.35$, which is also in excellent agreement with the values obtained from observations \citep[][]{fulton2018a,wu2019a}. For example, the dashed line in the right panel of \Cref{fig:comparison_prop_m} from observations by \citet{fulton2018a} corresponds to a slope of $\text{d log} R_p/ \text{d log} M_\ast \sim 0.35$ and \citet{wu2019a} reports a slope of $\text{d log} R_p/ \text{d log} M_\ast \simeq 0.24-0.35$ and finds that it extends to M-dwarfs of masses as low as $0.2 M_{\sun}$. We demonstrate in Section \ref{sec:results_stellar_prop} that this slope is, to first order, due to the dependence of the core-powered mass-loss mechanism on the bolometric luminosity of the host star as this dictates the outer boundary condition for atmospheric mass-loss.

\subsubsection*{Dependence on stellar mass as a function of orbital period and insolation flux}

\begin{figure*}
\centering
\includegraphics[width=1\textwidth,trim=280 470 130 310,clip]{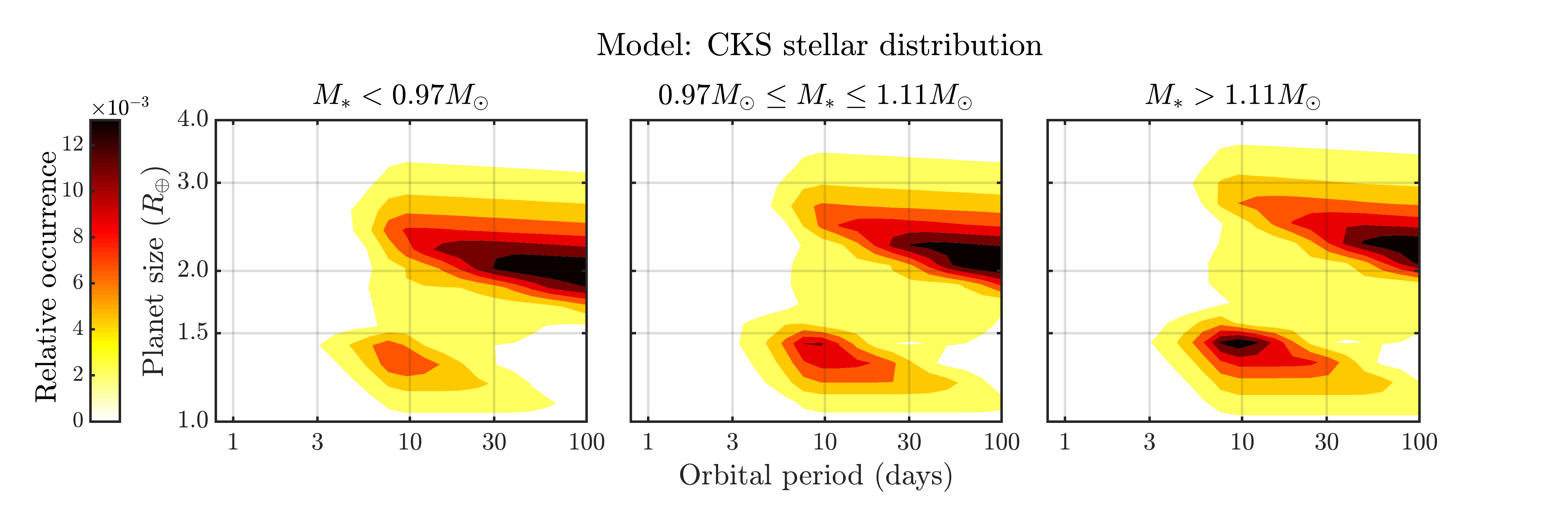}
\includegraphics[width=1\textwidth,trim=280 480 130 330,clip]{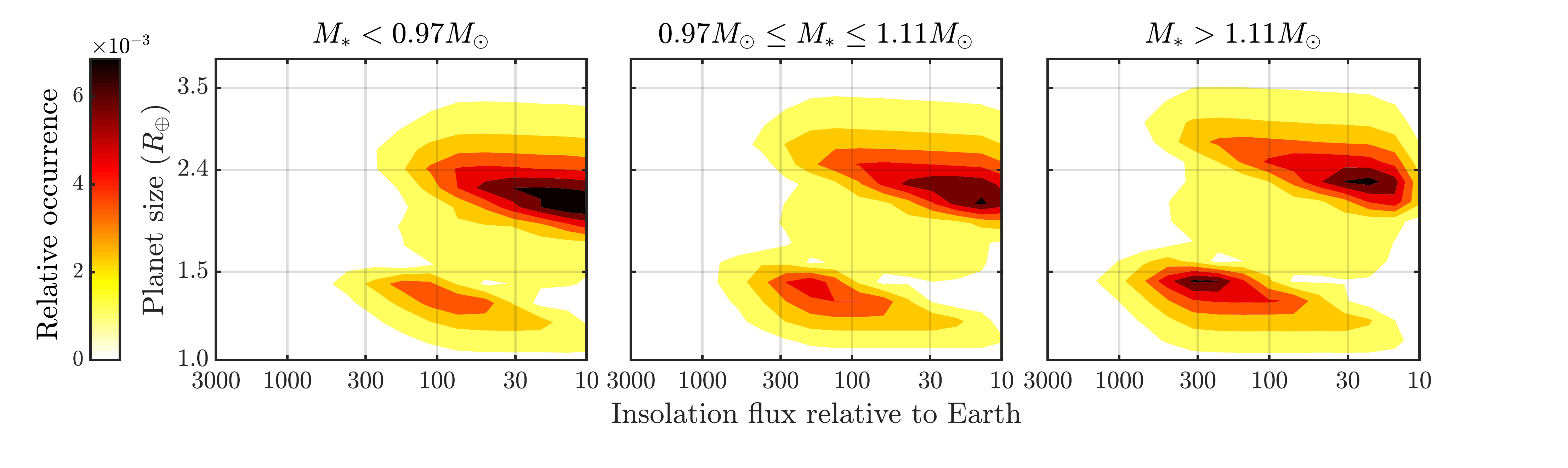}
\includegraphics[width=0.95\textwidth,trim=-25 0 0 0,clip]{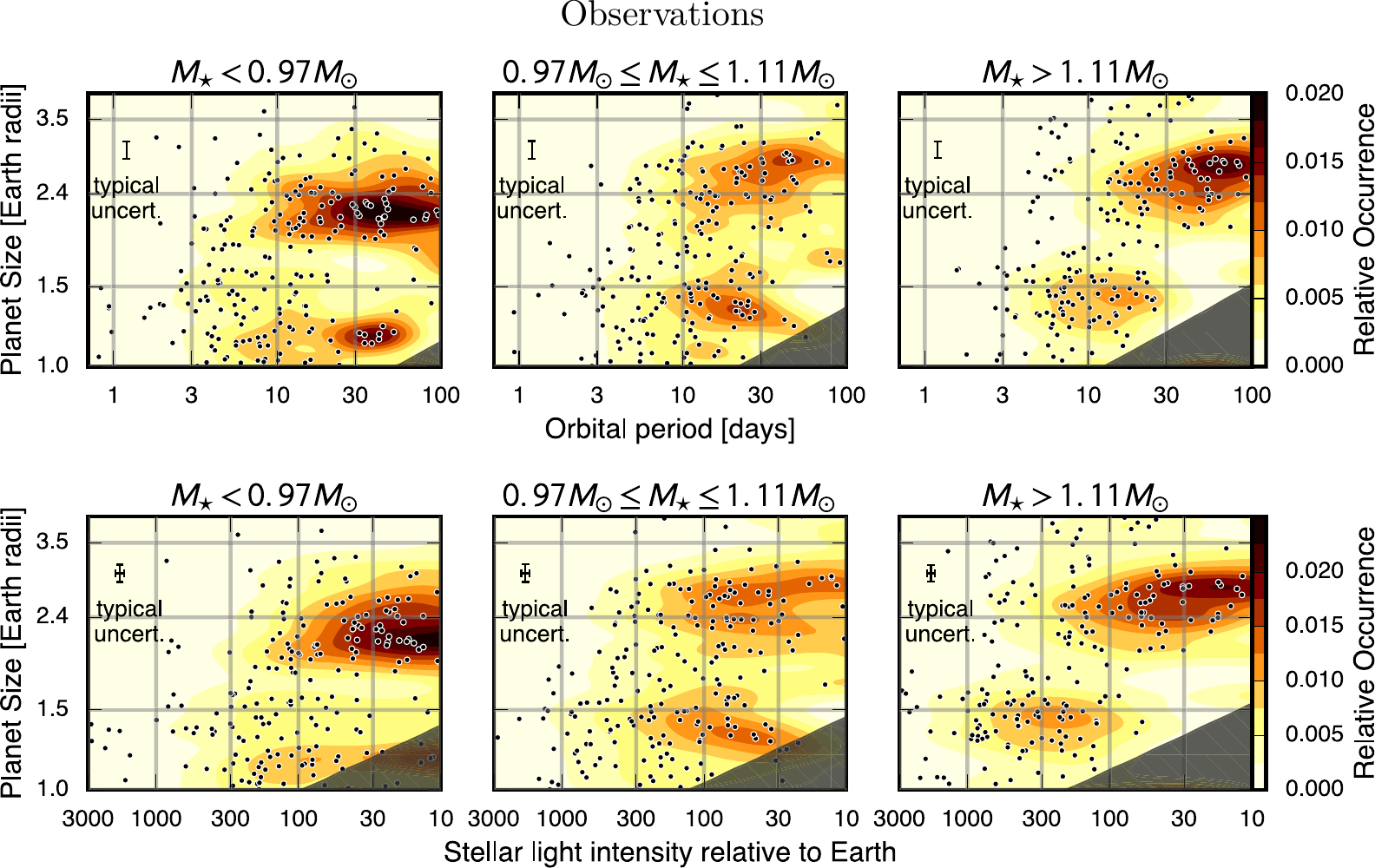} 
\caption{Comparison of core-powered mass-loss results (top two rows) with {completeness-corrected} observations (bottom two rows) from \citet[][reproduced with permission]{fulton2018a}. The planet size distribution is shown for three stellar mass bins as a function of orbital period (first and third row) and stellar insolation flux (second and fourth row). Our core-powered mass-loss results clearly show that the radius valley moves to larger planet sizes for more massive stars (see also \Cref{fig:comparison_prop_m}). In addition, we find that the super-Earth and sub-Neptune populations move to higher stellar insolation for more massive stars, which is due to the fact that the exoplanet period distribution is approximately independent of stellar mass.}
\label{fig:comparison_prop_m_bins}
\end{figure*}

Figure \ref{fig:comparison_prop_m_bins} displays the planet size distribution as a function of period (first row) and insolation (second row) for three stellar mass bins. The observations from \citet{fulton2018a} are shown for comparison in rows three and four. Our core-powered mass-loss results (displayed in the first and second row) clearly show that the radius valley moves to larger planet sizes for more massive stars as already shown in \Cref{fig:comparison_prop_m}. In addition, we find that the super-Earth and sub-Neptune populations move to higher stellar insolation for more massive stars. This is due to the fact that the exoplanet period distribution is, motivated by observations \citep[see][]{fulton2018a}, assumed to be independent of stellar mass. As a result, planets with the same orbital period move to higher insolation fluxes around higher mass, i.e. more luminous, stars. For a constant period distribution, insolation flux scales with stellar mass as $S\propto M_\ast^{\alpha-2/3}$, where $\alpha$ is the power-law index of the stellar mass-luminosity relation (see Section \ref{sec:results_stellar_prop_M} for details). Overall, we find very good agreement between the core-powered mass-loss results and observations. Based on the observational results shown in rows three and four of Figure  \ref{fig:comparison_prop_m_bins}, \citet{fulton2018a} reported that the planet size distribution shifts to higher stellar insolation flux around more massive stars and interpret this as a signature of photoevaporation. We question this interpretation based on the core-powered mass-loss results shown in rows one and two of the same figure. As discussed, the shift of the planet size distribution to higher stellar insolation flux around more massive stars is a direct result of the fact that the period distribution is approximately independent of stellar mass. In addition, one needs to be cautious with such claims until the effect of all stellar properties on the results are considered collectively since stellar age and metallicity conspire together such that sub-Neptunes are larger for younger and higher metallicity stars, both of which are correlated with stellar mass in the CKS dataset (see Section \ref{sec:results_stellar_prop} for details).

In addition, a shift in the exoplanet distribution to higher insolation as a function of stellar type has also been observed around M-dwarfs. \citet{hirano2018a} reported that around early, less XUV active M-dwarfs, planets are located at higher insolation flux in comparison to mid-to-late, more XUV active M-dwarfs; see their Figures 18 and 19. \citet{hirano2018a} interpret this as a signature of photoevaporation. However, if this is indeed a signature of photoevaporation transforming sub-Neptunes into super-Earths, then \citet{hirano2018a} should also have found rocky super-Earths at higher insolation around the more active M-dwarfs, as they are the remnants of the stripped sub-Neptunes. However, no such population is detected. Therefore, this observed shift in insolation is unlikely a signature of an atmospheric mass-loss mechanism. Instead, Figures 18 and 19 from \citet{hirano2018a} might be evidence of protoplanetary disk truncation around M-dwarfs which could be caused by  {disk-}photoevaporation, or some other process.

\subsection{Changes in planet size distribution with stellar metallicity}

\begin{figure*}
\centering
\includegraphics[width=.53\textwidth,trim=800 450 365 370,clip]{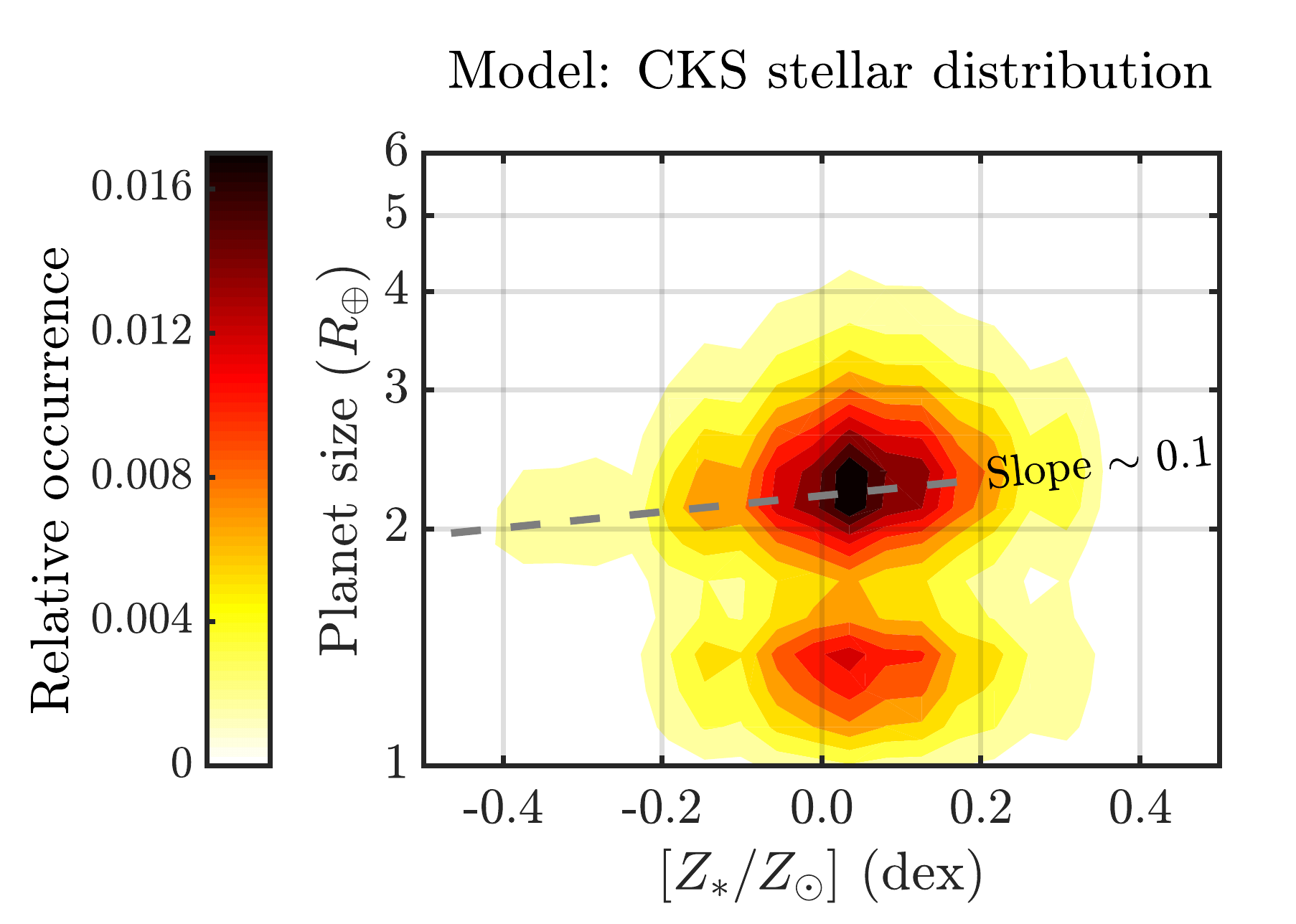}\hfill \includegraphics[width=.46\textwidth,trim=0 -5 0 0,clip]{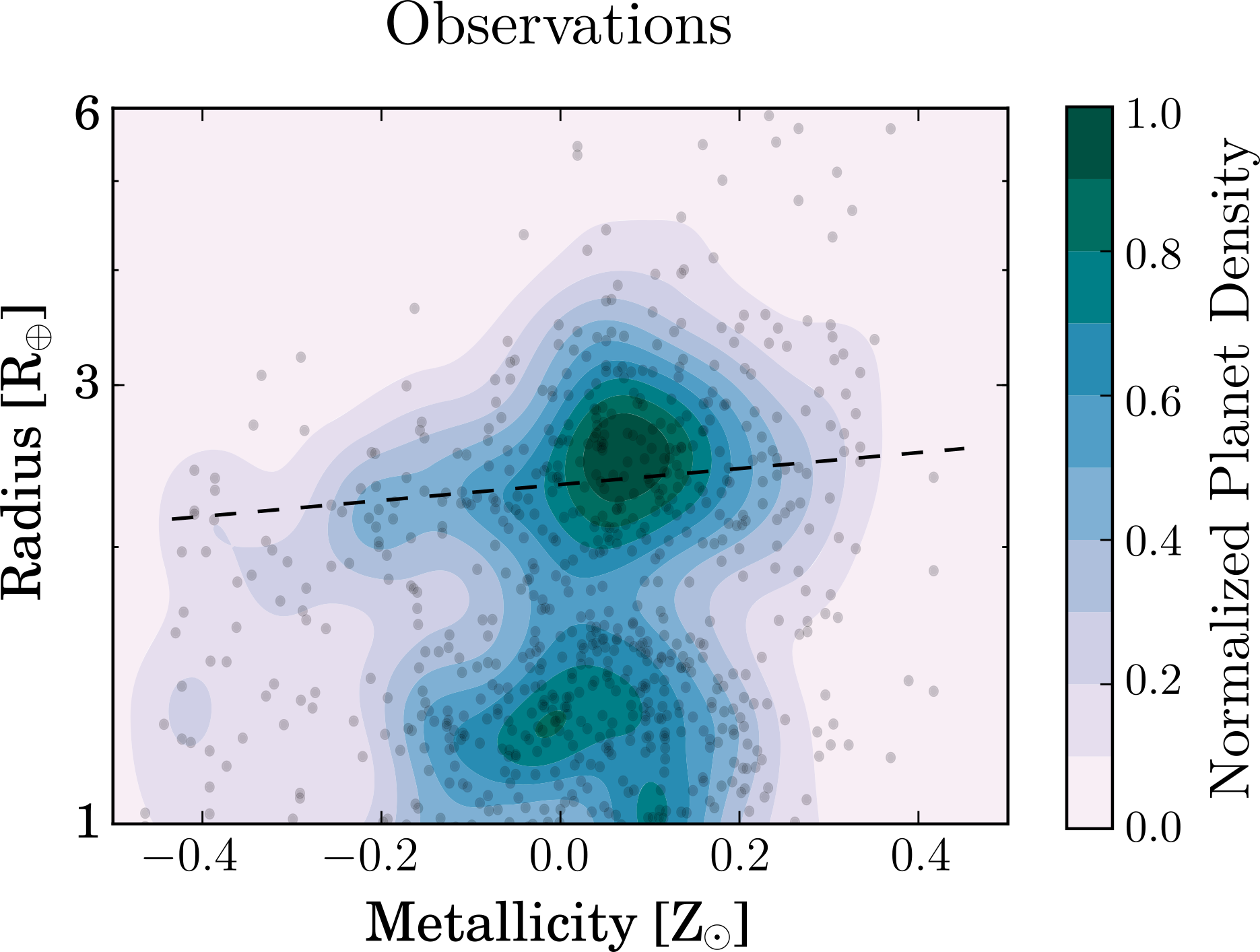} 
\caption{Comparison of core-powered mass-loss results as a function of metallicity, $[Z_\ast/Z_\odot ]$ (left panel), with {completeness-corrected} observations from \citet{owen2018a} (right panel). The dashed lines in both panels indicate the increase in the size of sub-Neptunes with metallicity with a slope of $\text{d log} R_p/ \text{d log} Z_\ast \sim 0.1$. Sub-Neptunes are on average larger around metal-rich stars, since higher atmospheric opacities result in longer thermal contraction and cooling timescales. In contrast, the location and slope of the radius valley, as well as the average sizes of super-Earths are, to first order, independent of metallicity; see Section \ref{sec:results_stellar_prop_Z} for details.}
\label{fig:comparison_prop_Z}
\end{figure*}

The left panel of \Cref{fig:comparison_prop_Z} shows core-powered mass-loss results as a function of stellar metallicity. We find that the size of sub-Neptunes increases with increasing stellar metallicity whereas neither the radius valley nor population of super-Earths show any visually discernible dependence on stellar metallicity. Specifically, we find that the average size of sub-Neptunes increases with stellar metallicity (represented by the inclination of the contours of the sub-Neptune population) such that $\text{d log} R_p/ \text{d log} Z_\ast \sim  0.1$. These finding are consistent with observations by \citet{petigura2018a}, \citet{dong2018a} and \citet{owen2018a}. The right panel of \Cref{fig:comparison_prop_Z} shows observations from  \citet{owen2018a} based on the CKS dataset \citep{johnson2017a,fulton2017a}. The dashed line, roughly drawn to indicate  the observational trend of planet size with stellar metallicity for the sub-Neptune population, has a slope of $\text{d log} R_p/ \text{d log} Z_\ast \sim 0.1$. We note here that  {observations show} hardly any change in the location of the valley as a function of stellar metallicity. This suggests that, to first order, changes in metallicity predominately impact the sizes of sub-Neptunes but not the super-Earth population. As we show in the next section (see Section \ref{sec:results_stellar_prop_Z}),  {these observations are consistent with results from our core-powered mass-loss model, which predicts no metallicity dependence of the radius valley and the super-Earth population. The observed planet size-metallicity trend displayed in the sub-Neptune population is also consistent with the results in Section \ref{sec:results_stellar_prop_Z}. However, as the trend in the sub-Neptune population is predominately due to that fact that the thermal contraction timescales are prolonged for higher envelop opacities/metallicities, the observed planet size-metallicity trend is not a unique signature of the core-powered mass-loss mechanism.}

\subsection{Changes in planet size distribution with stellar age}

\begin{figure*}
\centering
\includegraphics[width=0.52\textwidth,trim=795 450 380 370,clip]{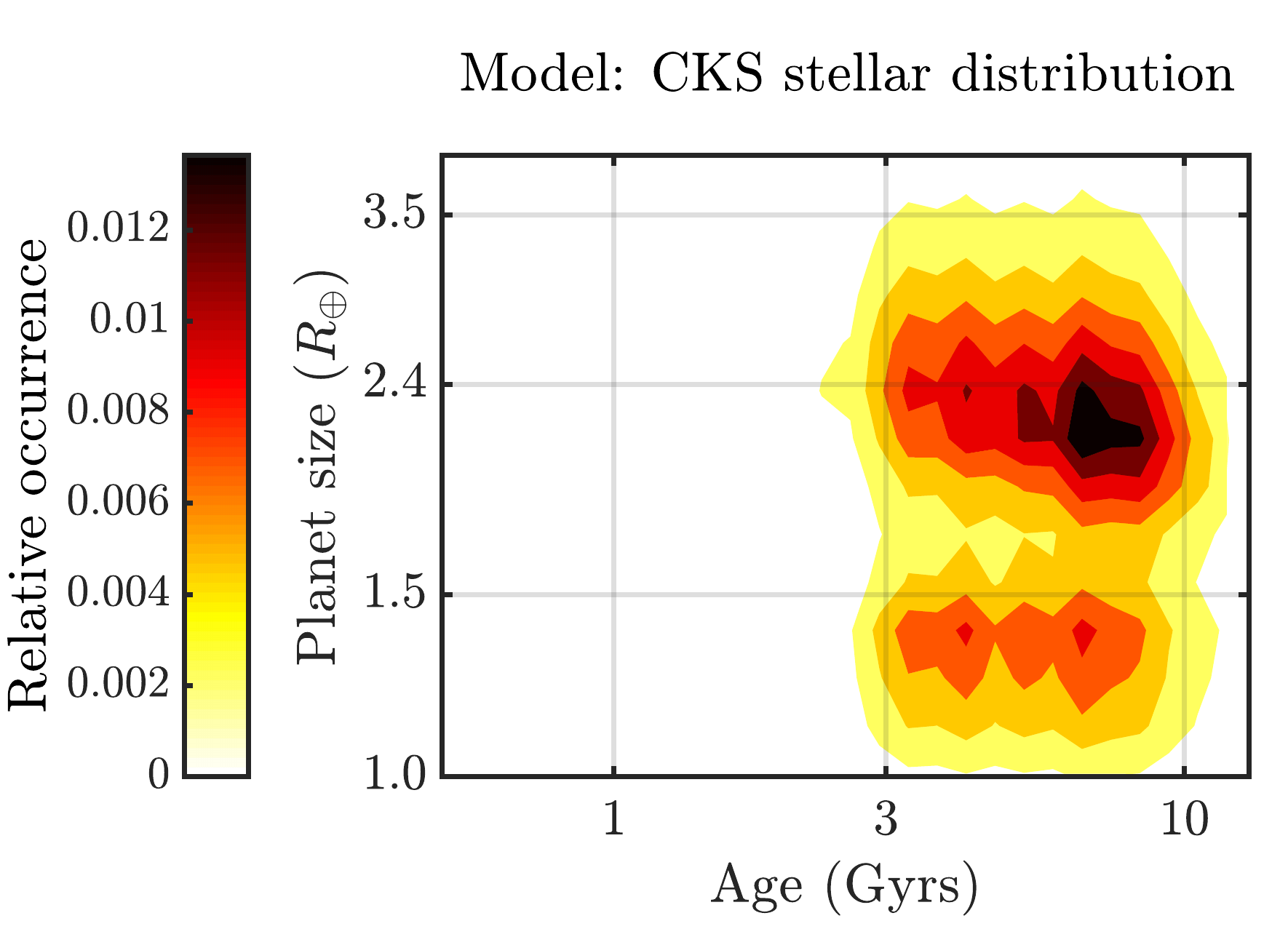} \hfill \includegraphics[width=0.45\textwidth,trim=380 450 850 370,clip]{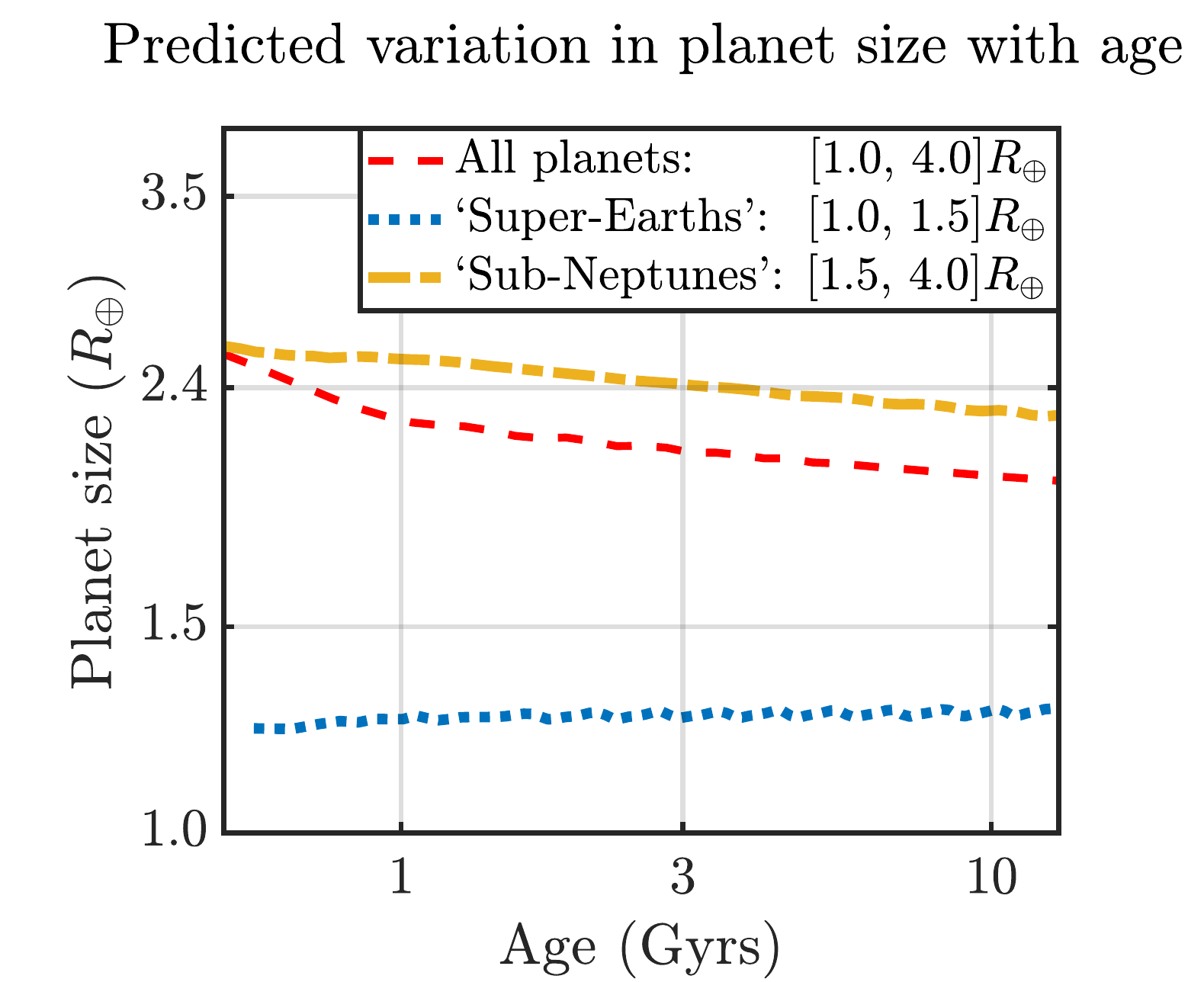} 
\caption{Core-powered mass-loss results as a function of age. The left panel shows theoretical results for a planet population around a distribution of host stars modeled after the CKS dataset \citep[][]{johnson2017a,fulton2017a}; the right panel shows our predicted average planet size with age. These results demonstrate that the average size of sub-Neptunes decreases significantly while the average size of super-Earths increases slightly with age. In addition, we predict that the relative occurrence of sub-Neptunes with respect to super-Earths decreases with age over Gyr timescales. These trends are due to the fact that typical mass-loss timescales are on the order of 0.5-1 Gyrs, which leads to an increase of super-Earths with respect to sub-Neptunes over these timescales. In addition the envelopes of sub-Neptunes will have had more time to cool and contract resulting in a decrease in the average sub-Neptune size with age; see Section \ref{sec:results_stellar_prop_A} and \Cref{fig:stellar_prop_var_A} for details.}
\label{fig:stellar_A_var}
\end{figure*}

\Cref{fig:stellar_A_var} shows core-powered mass-loss results vary as a function of age for a population of planets with host stars modeled after the CKS dataset \citep{johnson2017a,fulton2017a}. We  {predict} that sub-Neptunes are visibly smaller around older stars since their envelopes had more time for thermal contraction in older systems, whereas super-Earths have similar but slightly larger sizes as a function of age.  {We note here, no observations are shown in \Cref{fig:stellar_A_var}}. Given the small range in stellar ages and their relatively larger uncertainties, no direct statistical comparison between core-powered mass-loss results and observations is currently possible (but see discussion in \Cref{sec:results_stellar_prop_A} for details).
\\

{In this section, we  {discussed} our core-powered mass-loss results for a population of planets with host stars modeled after the CKS dataset and showed that the core-powered mass-loss mechanism can reproduce and explain a multitude of observational trends \citep[e.g.,][]{fulton2018a,owen2018a,petigura2018a,dong2018a,wu2019a}. {In the next section, we explain the key physical properties of the core-powered mass-loss mechanism that lead to the results presented here and make observational predictions for the core-powered mass-loss mechanism as a function of stellar properties.}}

\section{Results: Analytical scalings \& Dependence on stellar properties}\label{sec:results_stellar_prop}

In this section, we explore and explain, analytically and numerically, the dependence of the core-powered mass-loss mechanism on stellar mass, metallicity and age by only varying one of these stellar parameters at a time. 

We assume{, with the exception of Section \ref{sec:results_planet_stellar_mass},} the same planet mass and orbital period distribution for all host stars such that any changes in the resulting planet size distribution can be unambiguously attributed to changes in the host star properties. Unlike in Section \ref{sec:results_comparison_w_obs}, where we modeled the host stars after the CKS dataset, here we assume a uniform distribution for the single stellar parameter that we are varying. For the parameters that are held constant, we assume solar values for metallicity and mass, and an age of 3 Gyrs.

\begin{figure*}
\centering
\includegraphics[width=0.5\textwidth,trim=305 450 890 365,clip]{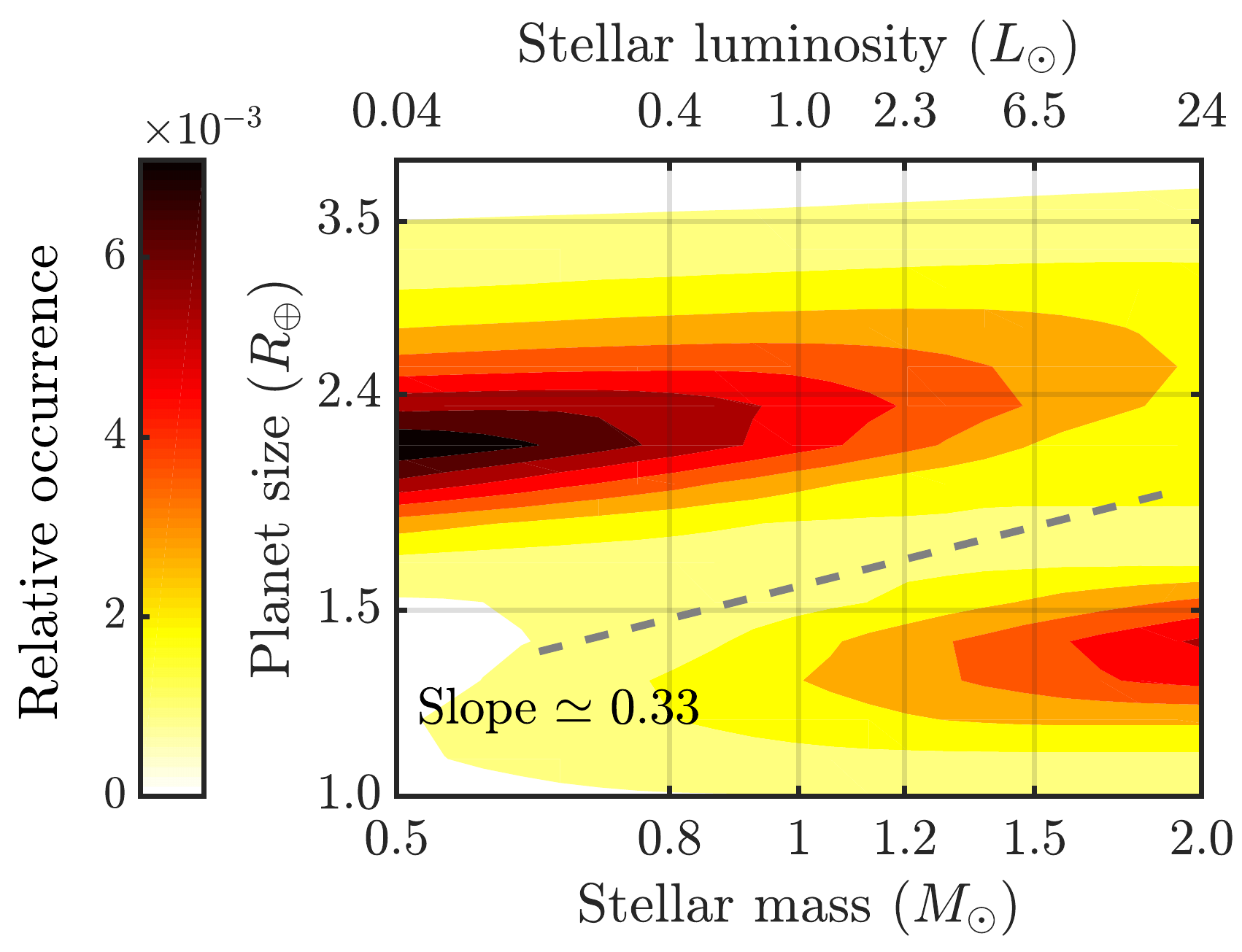} 
\includegraphics[width=0.49\textwidth,trim=390 315 900 600,clip]{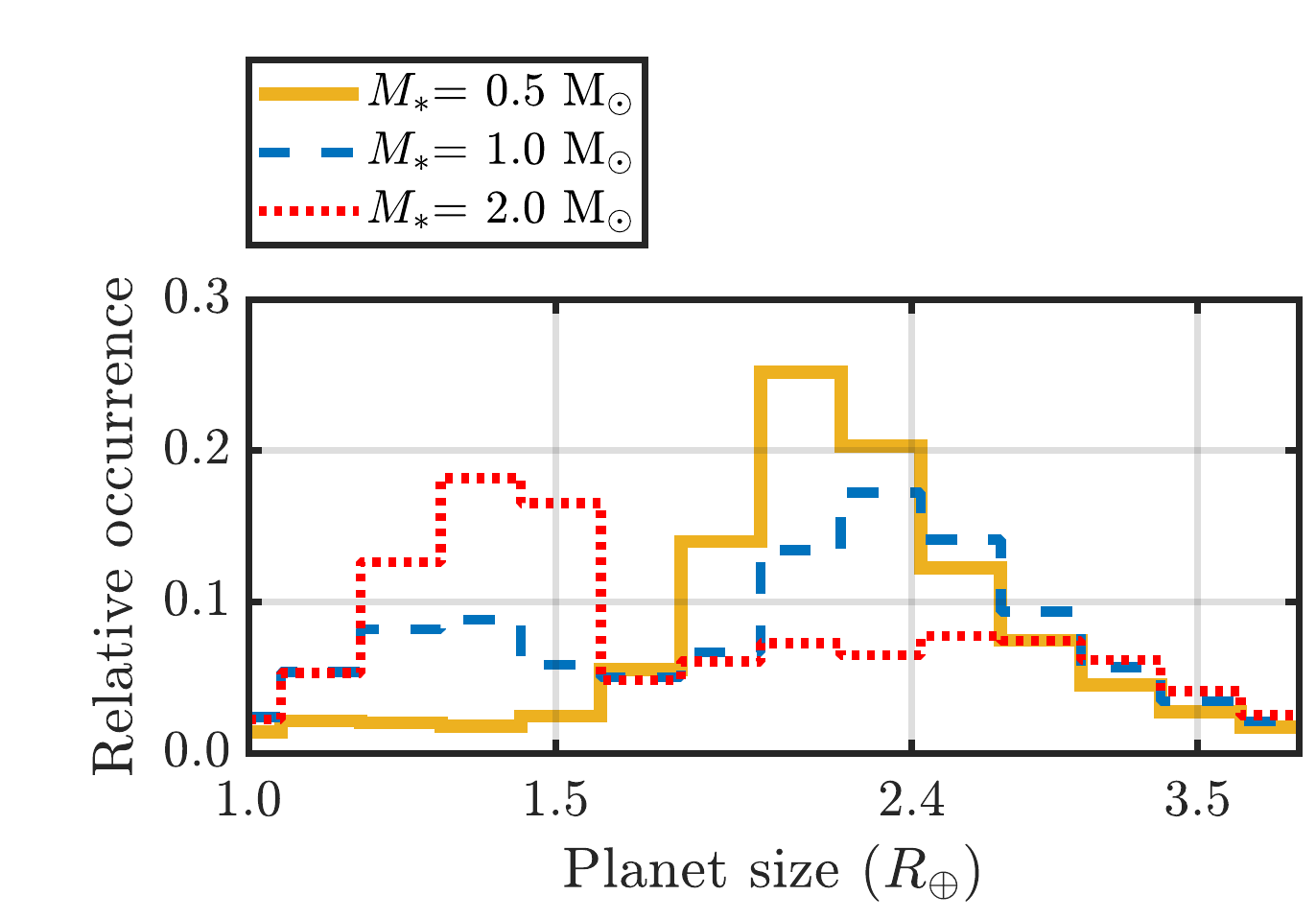}
\caption{Core-powered mass-loss results as a function of stellar mass (luminosity) for planets around solar metallicity stars after 3 Gyrs of thermal evolution and mass-loss. Here, only the stellar mass is varied while keeping all other parameters the same.  The left panel shows the two-dimensional distribution of planet size as a function of stellar mass (luminosity); the right panel displays the corresponding histograms of planet sizes for stellar masses of $0.5 M_{\sun}$ (yellow-solid line), $1M_{\sun}$ (blue-dashed line), and $2M_{\sun}$ (red-dotted line), respectively. The dashed line in the left panel corresponds to the slope of the radius valley, ${\text{d log}R_p}/{\text{d log} M_\ast} \simeq 0.33$, derived analytically in \Cref{eq:slope_R_p_M_s}.}
\label{fig:stellar_prop_var_M}
\end{figure*}

\subsection{Dependence on stellar mass} \label{sec:results_stellar_prop_M}
Stellar mass plays a role when modelling atmospheric loss driven by the core-powered mass-loss mechanism in two ways: (i) it relates the orbital period distribution of the planet population to the semi-major axis and (ii) it dictates the equilibrium temperature $T_{\text{eq}}$ for each planet given its distance from the host star. The equilibrium temperature depends on the bolometric luminosity $L_\ast$ and the semi-major axis $a$ such that $T_{\text{eq}} \propto L_\ast^{1/4} a^{-1/2}$. The bolometric luminosity, in turn, can be expressed as $L_\ast \propto T_{\text{eff}}^4 R_\ast^{-2}$, where $T_{\text{eff}}$ is the effective temperature of the star and $R_\ast$ the stellar radius. The CKS dataset provides both, stellar radii and effective temperatures. However, the bolometric luminosity is also strongly correlated with the stellar mass $M_\ast$ such that
\begin{equation}
L_\ast/L_\odot = (M_\ast/M_\odot)^\alpha.
\end{equation}
The dependence of the core-powered mass-loss mechanism on stellar mass is, to first order, driven by the stellar mass-luminosity relation, since more massive stars have higher luminosity which leads to higher $T_{\text{eq}}$ and hence more mass loss from the planet.

While we directly use stellar radius and effective temperature to derive a star's luminosity for the numerical simulation results discussed in Section \ref{sec:results_comparison_w_obs}, in this section we use the stellar mass-luminosity relation to isolate the effect of stellar mass on the resulting radius valley and planet size distribution. Given the stellar mass-luminosity relation, for a fixed period distribution we get $T_{\text{eq}} \propto M_\ast^{(\alpha/4) - (1/6)}$, which sets the temperature at a planet's Bondi radius. \citet{gupta2019a} showed that the slope and location of the valley in the planet distribution is set by the condition $t_{loss}=t_{cool}$. Furthermore, they showed that this condition is very well approximated by $ G M_p/c_s^2 R_{rcb} =$ constant, due to the exponential dependence of $t_{loss}^B$ (see Section \ref{sec:model}). Substituting for the speed of sound, the mass-radius relation of the core, and using the fact that $R_{rcb}= R_p \simeq  2R_c$ yields
\begin{equation}\label{eq:exp_slope}
R_p^3 T_{\text{eq}}^{-1} \simeq \text{constant}.
\end{equation}
Substituting for $T_{\text{eq}}$ using the stellar mass-luminosity relation we find
\begin{align}\label{eq:slope_R_p_M_s}
    \frac{\text{d log}R_p}{\text{d log}M_\ast} = \frac{3 \alpha - 2}{36} =0.33 \text{ for } \alpha=4.6.
\end{align}
{Here we evaluated the last equality for $\alpha = 4.6$, which is the best-fit value we measured from the CKS dataset.}
This analytical estimate of the slope is in good agreement with the observations based on the CKS dataset which find ${\text{d log}R_p}/{\text{d log} M_\ast} \sim 0.35$ \citep[e.g.,][]{fulton2018a,wu2019a}, with the results of our numerical simulations discussed in Section \ref{sec:results_comparison_w_obs} and with the numerical results {based on the mass-luminosity relation, which is only used in \Cref{sec:results_stellar_prop_M},} as shown in \Cref{fig:stellar_prop_var_M}.
In addition, we predict that, if planetary evolution is dominated by core-powered mass-loss, the radius valley will have a slope $\text{d log}R_p/\text{d log}M_\ast = (3 \alpha - 2)/36 $ and hence will be steeper for planets around host star  {populations} with larger $\alpha$ and vice-versa \citep[e.g.,][]{eker2018a}. This trend would be weakened if photoevaporation or other mass-loss processes also play a significant role in shaping the exoplanet radius distribution.

{We note here, while the value of $\alpha$ that we measure directly from the CKS dataset is somewhat larger than what is predicted by simple theoretical models, it is in agreement with observational measurements of the mass-luminosity relation for stars in the solar neighbourhood \citep{eker2018a}. For example, \citet{eker2018a} find values of $\alpha$ ranging from 4.3 to 5.7 for the relevant stellar masses in the CKS dataset. Furthermore, the CKS dataset is a magnitude-limited rather than a volume-limited sample thereby including more high luminosity/mass stars.}

\Cref{fig:stellar_prop_var_M} displays the planet size distribution as a function of stellar mass (luminosity) assuming that all host stars are 3 Gyrs old and have solar metallicity. The left panel of \Cref{fig:stellar_prop_var_M} shows the location of the valley and that the size of super-Earths and sub-Neptunes increases with increasing stellar mass (or luminosity) as derived analytically in \Cref{eq:slope_R_p_M_s} and as shown by the black-dashed line in the figure. In other words, planets around more massive stars are, on average, bigger in size. In addition, the relative abundance of sub-Neptunes decreases and that of super-Earths increases with increasing stellar mass. This is because planets around more massive or luminous stars are more susceptible to losing their atmospheres as they have higher equilibrium temperatures which result in shorter mass-loss timescales and thus a higher fraction of sub-Neptunes become super-Earths around more massive stars. These two results are also evident from the right panel of \Cref{fig:stellar_prop_var_M} which shows the radius distribution for host star masses corresponding to $0.5 M_{\sun}$, $1.0 M_{\sun}$ and $2.0 M_{\sun}$, respectively. These results explain observations reported by \citet{fulton2018a} and \citet{wu2019a} who find that the radius valley increases in planet size with increasing stellar mass.

Assuming that the orbital period distribution is independent of stellar mass, this observation seems surprising under a basic framework of photoevaporation since the XUV flux roughly remains constant as a function of stellar mass and hence should yield an approximately flat radius valley as a function of host star mass; see also \citet{wu2019a}. This is because the ratio of XUV to bolometric flux decreases with stellar mass \citep[e.g.,][]{jackson2012a,tu2015a} while the bolometric flux increases. These two trends roughly cancel each other \citep{owen2018a}. Therefore, in order to explain radius valley's dependence on stellar mass with photoevaporation models, \citet{wu2019a} invoked a linear correlation between the planet mass distribution and host star mass  {and thereby is able to match the observations}.

As shown above and in Section \ref{sec:results_comparison_w_obs}, the core-powered mass-loss mechanism yields a shift in the radius valley to larger planet sizes around more massive stars with a slope given by $\text{d log} R_p/\text{d log} M_\ast \simeq 0.33$, in agreement with observations. We therefore find,  {in contrast} to photoevaporation models, no evidence for a  {linear} correlation between planet and stellar mass. We nonetheless investigate the imprint that a linear correlation between planet and stellar mass would have on the observed planet size distribution in the context of the core-powered mass-loss mechanism in Section \ref{sec:results_planet_stellar_mass}.

\begin{figure*}
\centering
\includegraphics[width=0.5\textwidth,trim=305 450 890 407,clip]{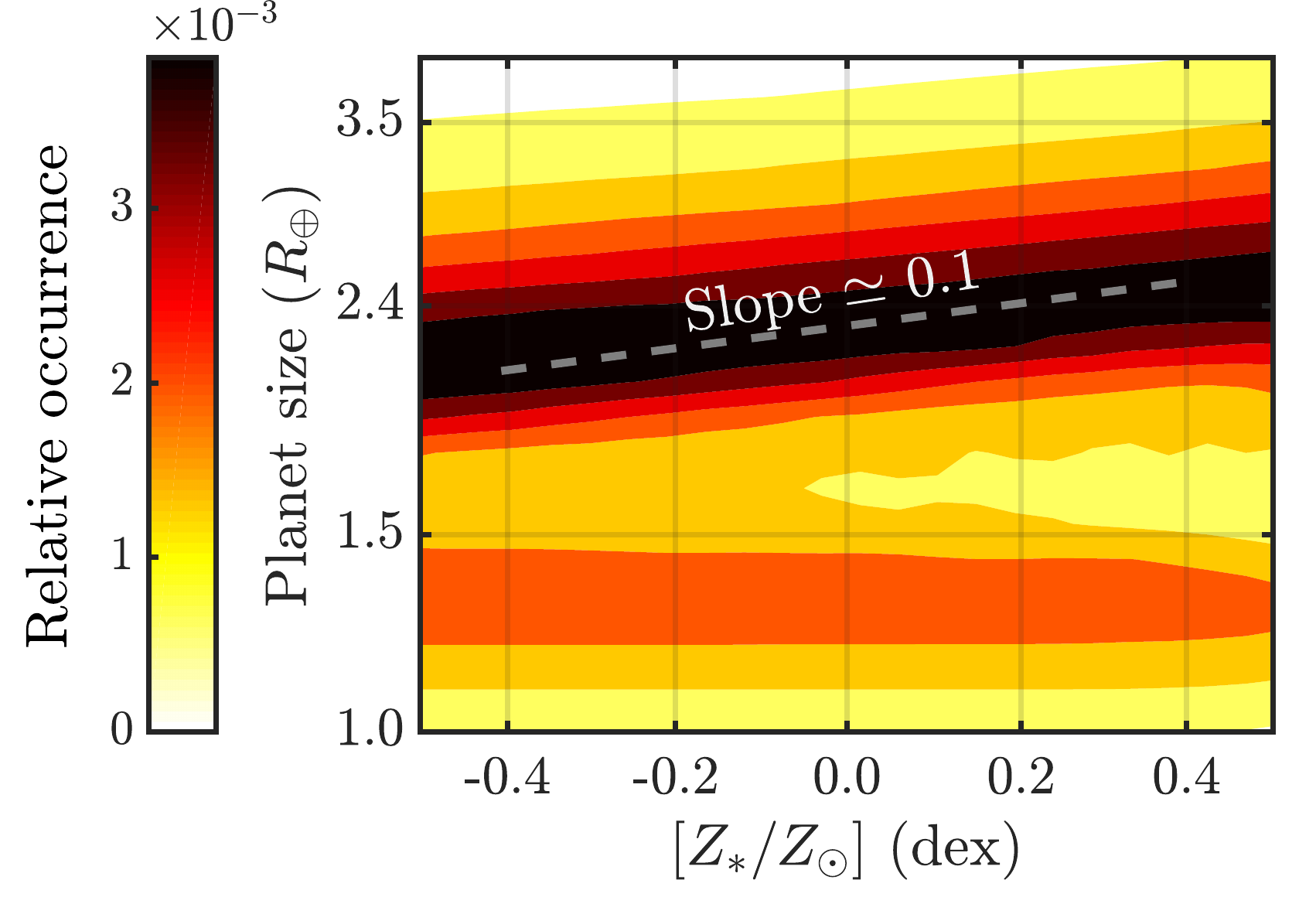} 
\includegraphics[width=0.49\textwidth,trim=390 315 900 600,clip]{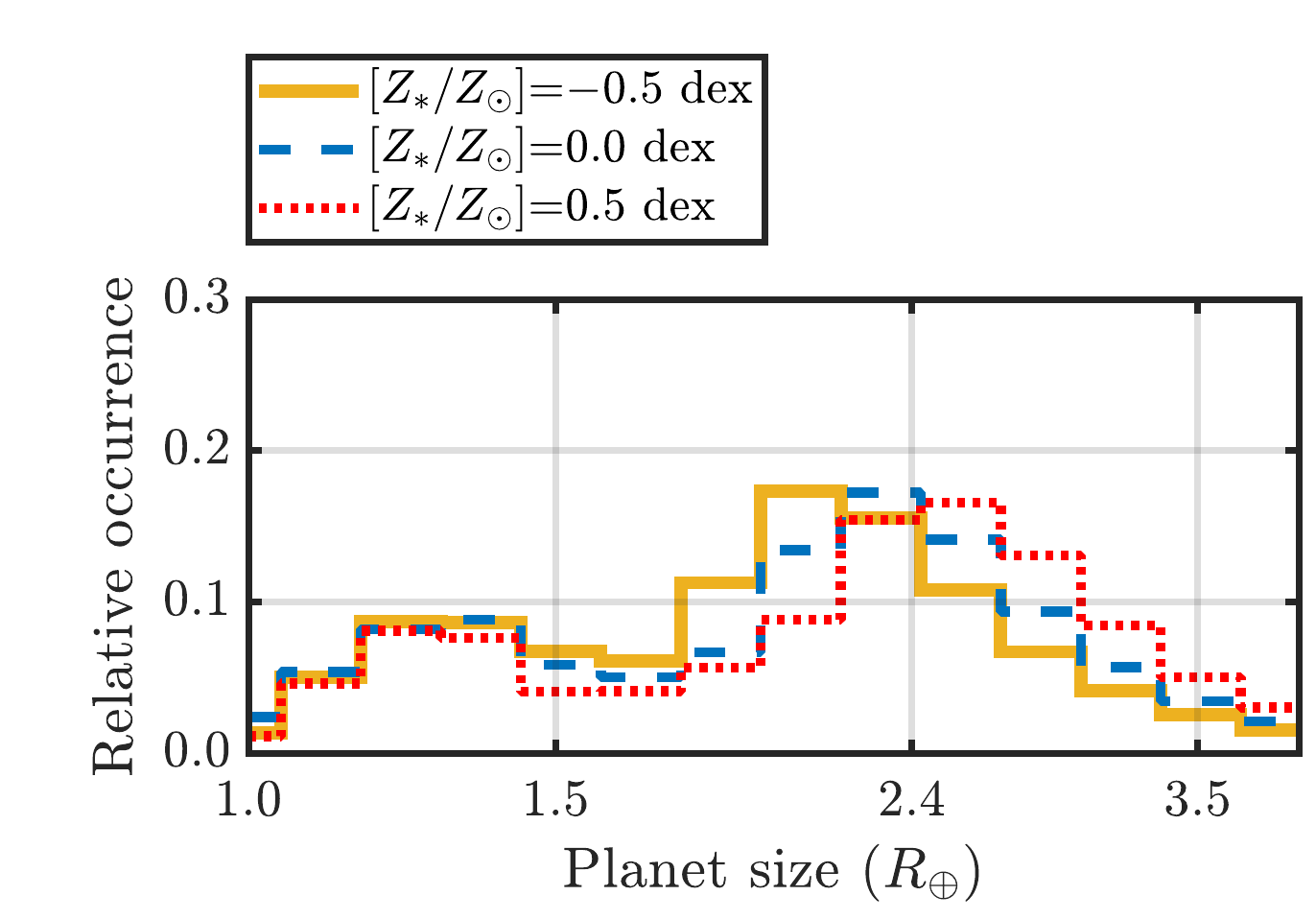}
\caption{Core-powered mass-loss results as a function of stellar metallicity for planets around solar mass stars after 3 Gyrs of thermal evolution and mass-loss. The left panel shows the distribution of planet size as a function of metallicity $\equiv [ Z_\ast / Z_\odot ]$; the right panel displays the corresponding histograms of planet size for metallicities of $-0.5$ dex (yellow-solid line), $0.0$ dex (blue-dashed line), and $0.5$ dex (red-dotted line), respectively. The gray-dashed line in the left-panel shows the increase in the size of sub-Neptunes with metallicity, i.e., ${\text{d log}R_p}/{\text{d log} Z_\ast} \simeq 0.1$, derived analytically in \Cref{eq:slope_z}. In contrast to the sub-Neptune population, the location of the valley, specifically the lower edge of the valley defining the upper envelope of the super-Earth population, shows no significant dependence on metallicity. Furthermore, this figure also shows that the relative occurrence of sub-Neptunes with respect to super-Earths increases with increasing metallicity.}
\label{fig:stellar_prop_var_Z}
\end{figure*}

\subsection{Dependence on stellar metallicity}\label{sec:results_stellar_prop_Z}
The rate at which planets cool and contract  depends on the opacity of the envelope,  $\kappa$, as this sets the rate of radiative diffusion through the radiative-convective boundary (see \Cref{eq:L,eq:t_cool}). If the opacity at the radiative-convective boundary is proportional to the stellar metallicity, $Z_\ast$, as we assume in Section \ref{sec:model}, then the planetary cooling timescale is directly proportional to the stellar metallicity such that $t_{cool}\propto Z_\ast$. In other words, a planet around a metal-poor star, i.e., a planet with lower atmospheric opacity, loses its energy on a shorter timescale than around a higher metallicity star. As a result, sub-Neptunes will be larger around higher metallicity stars than metal poor stars when comparing systems of the same age. We can derive an analytical estimate for the size of sub-Neptunes as a function of metallicity by assuming that atmospheric mass-loss can be neglected such that $M_{atm}=$ constant and thus cooling and contraction due to mass-loss can be ignored. In this case, \Cref{eq:f,eq:M_atm,eq:E_cool,eq:L,eq:opacity} can be combined to yield
\begin{align}\label{eq:tzR}
    t \propto Z_\ast \Delta R^{-\frac{\beta \gamma +1}{\gamma-1}},
\end{align}
where $t$ is the age of the system and where we used \Cref{eq:f} and the assumption that $M_{atm}=$ constant to eliminate the dependence on the density at the $R_{rcb}$. Hence, from \Cref{eq:tzR} we have that at a given time $\Delta R \propto Z_\ast^{(\gamma-1)/(\beta \gamma +1)}$ such that
\begin{align}\label{eq:slope_z}
    \frac{\text{d log}R_p}{\text{d log} Z_\ast} &= \frac{\Delta R}{R_p} \left(\frac{\gamma-1}{\beta \gamma +1} \right) \simeq \frac{5}{46},
\end{align}
where we substitute $\gamma=7/5$, $\beta=0.6$ (see Equation \ref{eq:opacity}) and $\Delta R/R_p \simeq 1/2$ in the last step.
We note here that approximating $\Delta R/R_p$ as $\simeq 1/2$, which is essentially our initial condition, results in the maximum slope possible and since $\Delta R/R_p$ decreases over time as the envelope cools and contracts. The value calculated in Equation (\ref{eq:slope_z}) is therefore an overestimate of the slope for a planet population with an age of 3 Gyrs.

The left panel of \Cref{fig:stellar_prop_var_Z} shows the planet size distribution as a function of metallicity, where we assume that the planet's opacity scales linearly with stellar metallicity (see \Cref{eq:opacity}). {The size trend with metallicity and slope of the sub-Neptune population from our numerically evolved planet population agrees well with the simple analytical estimate derived in Equation (\ref{eq:slope_z}) above.} In addition, this analytical scaling matches the observations \citep{owen2018a} and our numerical results discussed in Section \ref{sec:results_comparison_w_obs} where ${\text{d log}R_p}/{\text{d log} Z_\ast} \sim 0.1$; see also \Cref{fig:comparison_prop_Z}. Our results are also consistent with observations reported by \citet{dong2018a}, \citet{owen2018a} and \citet{petigura2018a} who all find that sub-Neptunes are larger around more metal-rich stars. Our results similarly apply to M-dwarfs and can explain observations that find that planets, on average, are larger around higher metallicity M-dwarfs \citep{hirano2018a}. These results are also evident from the right panel of \Cref{fig:stellar_prop_var_Z} which shows the planet size distribution for host star metallicities of $-0.5$, $0.0$ and $0.5$ dex, respectively.

In contrast to the sub-Neptune population, core-powered mass-loss predicts a negligible dependence of the location of the radius valley and sizes of super-Earths as a function of metallicity. This is because, as shown in \citet{gupta2019a} and summarized in Section \ref{sec:results_stellar_prop_M}, the slope of the radius valley is, to first order, given by $ G M_p/c_s^2 R_{rcb} =$ constant. Therefore, as long as the spontaneous mass-loss phase (boil-off phase) results in initial planet sizes $R_{rcb}= R_p \simeq  2R_c$ independent of envelope metallicity, we find that the slope of the valley, especially the lower edge that marks the upper envelope of the super-Earth population, does not depend on metallicity. This explains the almost flat contours of the super-Earth population shown in \Cref{fig:stellar_prop_var_Z} and the flat valley shown in \Cref{fig:comparison_prop_Z} and is the reason why, to first order, only the sizes of sub-Neptunes depend on stellar metallicity but not the location of the valley or the sizes of super-Earths.

In reality, there is a weak dependence of the radius valley and super-Earth sizes on metallicity because the atmospheric mass-loss timescales increase with stellar metallicity such that at a given time, planets with higher atmospheric opacities will have lost less of their envelopes than those with lower envelope opacities. This leads, for a given age, to smaller cores (super-Earths) around metal-rich stars compared to metal poor ones. This in addition implies that, at a given time, the relative abundance of super-Earths to sub-Neptunes should be higher around lower metallicity stars, as can be seen in \Cref{fig:stellar_prop_var_Z}. These results offer an explanation as to why the relative occurrence of sub-Neptunes increases with increasing metallicity as found observationally by \citet{petigura2018a} and \citet{dong2018a}.

\begin{figure*}
\centering
\includegraphics[width=0.5\textwidth,trim=305 450 890 410,clip]{A_sVsR_p_new.pdf} 
\includegraphics[width=0.49\textwidth,trim=390 315 900 600,clip]{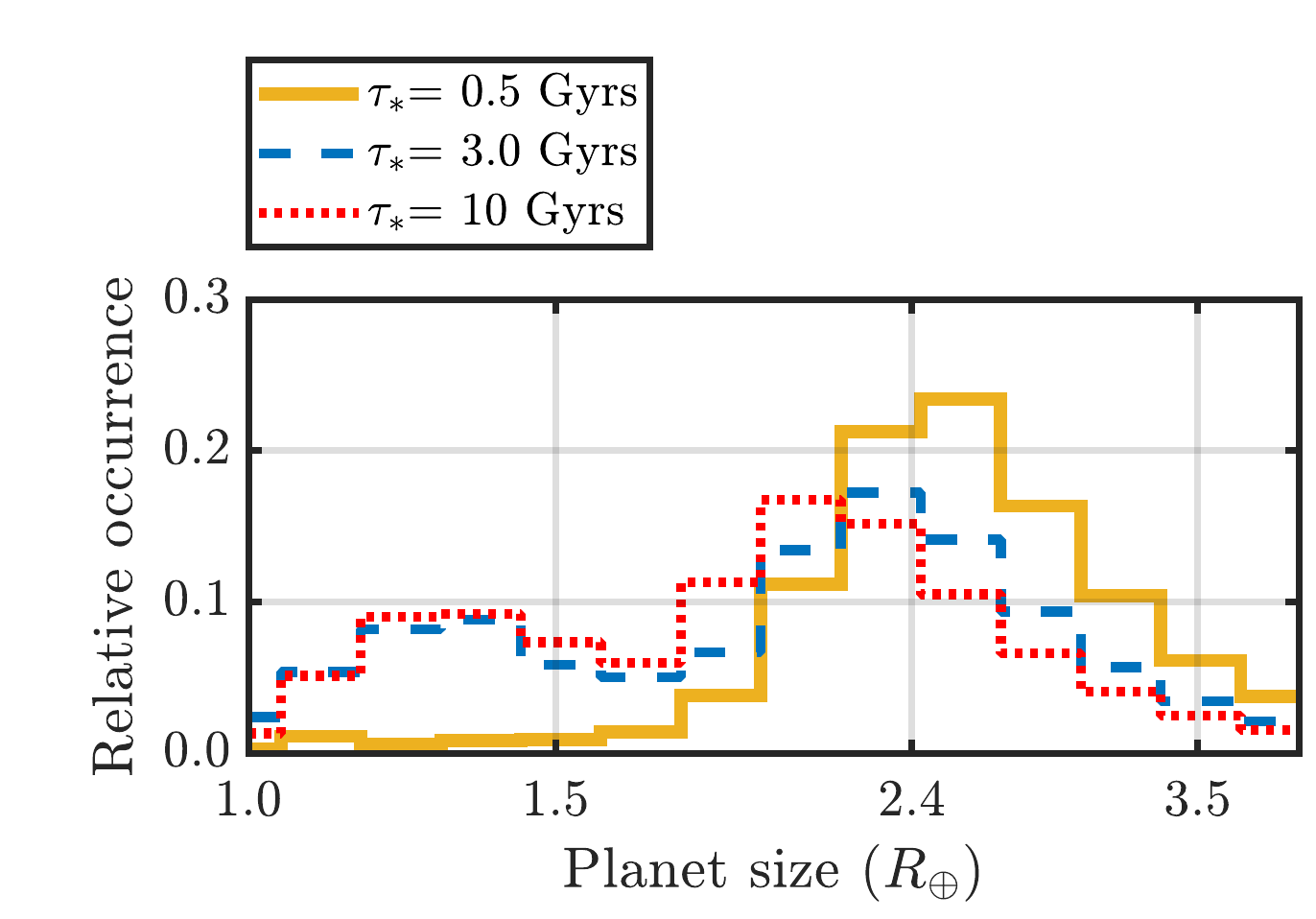}
\caption{Core-powered mass-loss results as a function of stellar age for planets with solar mass and solar metallicity host stars. The left panel shows the distribution of planet sizes as a function of age; the right panel displays the corresponding histograms of planet size for ages of $0.5$ Gyrs (yellow-solid line), $3.0$ Gyrs (blue-dashed line), and $10$ Gyrs (red-dotted line), respectively. The dashed line in the left panel shows the decrease in the size of sub-Neptunes with age, i.e., ${\text{d log}R_p}/{\text{d log} \tau_\ast} \simeq -0.1$, derived analytically in \Cref{eq:slope_t}. In contrast to the sub-Neptunes, the average size of super-Earths increases with age. In addition, this figure shows that the relative abundance of super-Earths to sub-Neptunes increases over time, as expected.}
\label{fig:stellar_prop_var_A}
\end{figure*}

\subsection{Dependence on stellar age}\label{sec:results_stellar_prop_A}

Atmospheric mass-loss driven by the cooling of the underlying core depends on the age of the system as mass-loss and thermal evolution extend, in contrast to photoevaporation, over Gyr timescales (see Section \ref{sec:model} for details). This implies that sub-Neptunes around older stars will be smaller than those around younger stars since the former will have had more time to cool and contract. Moreover, planets that eventually become super-Earths have had more time to complete their mass loss in older systems. Thus, older stars will have a higher abundance of super-Earths relative to sub-Neptunes compared to younger stars. Specifically, we expect a drastic change in the planet size distribution between stars that are younger and older than the typical core-powered mass-loss timescale, which is of the order of a Gyr. Due to these long mass-loss timescales, we predict the transformation of sub-Neptunes into super-Earths to continue over Gyr timescales.

Assuming a constant metallicity, we have from \Cref{eq:tzR} that the radii of the sub-Neptune population vary with age as
\begin{align}\label{eq:slope_t}
    \frac{\text{d log}R_p}{\text{d log} \tau_\ast} &= -\frac{\Delta R}{R_p} \left(\frac{\gamma-1}{\beta \gamma +1} \right) \simeq -\frac{5}{46},
\end{align}
where we again substitute $\gamma=7/5$, $\beta=0.6$ and $\Delta R/R_p \simeq 1/2$ in the last step. Just like for the metallicity dependence, the value calculated in Equation (\ref{eq:slope_t}) is an overestimate of the slope for a planet population with an age of 3 Gyrs (see discussion following Equation (\ref{eq:slope_z}) for details). Nonetheless, this analytical estimate agrees with our numerical results shown in \Cref{fig:stellar_prop_var_A}.

The left panel of \Cref{fig:stellar_prop_var_A} shows the changes in planet size as a function of age for solar mass and solar metallicity host stars. As expected, sub-Neptunes decrease in size with increasing stellar age as their envelopes cool and contract. Moreover, we find an increase in the average size of super-Earths with age as larger (more massive) cores are being stripped of their envelopes over time. This, in addition, leads to an increase in the number of super-Earths relative to sub-Neptunes with time. These results are displayed in the right panel of \Cref{fig:stellar_prop_var_A} which shows the planet size distribution for host stars with ages of $0.5$, $3.0$ and $10.0$ Gyrs. The core-powered mass-loss mechanism therefore predicts that sub-Neptunes decrease in size, while the average size of super-Earths increases with stellar age and that the relative abundance of super-Earths with respect to sub-Neptunes increases with time, especially over periods from 500 Myrs to 3 Gyrs.

The relatively large uncertainties in most stellar ages currently do not allow for a detailed comparison between predictions from the core-powered mass-loss mechanism and observations. However, according to \citet{mann2016a} and other ZEIT studies, younger planets (in stellar clusters $\sim$ 10-650 Myrs old) are on average bigger than the older \textit{Kepler} planets ($\gtrsim$ 1 Gyr old); see Figure 12 in \citet{rizzuto2018a}. According to these studies, close-in planets likely lose significant portions of their atmospheres even after the first few hundred million years. If correct, these observations directly confirm predictions from  core-powered mass-loss with its Gyr-long mass-loss timescales \citep[see also][]{ginzburg2016a} and would be challenging to explain with photoevaporation which mostly occurs over the first 100 Myrs.

\begin{figure*}
\centering
\includegraphics[width=.33\textwidth,trim=825 450 385 370,clip]{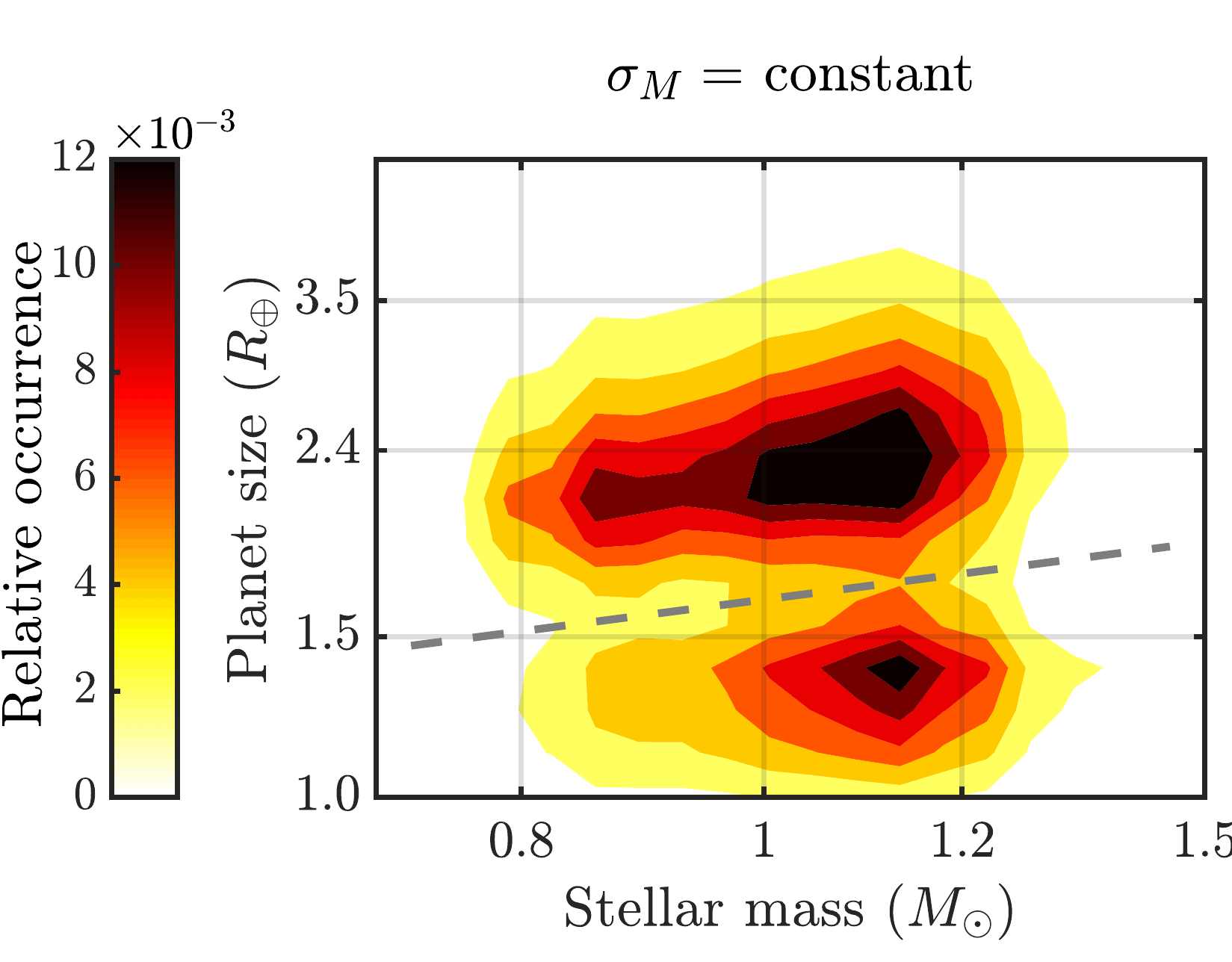} 
\includegraphics[width=.33\textwidth,trim=825 450 385 370,clip]{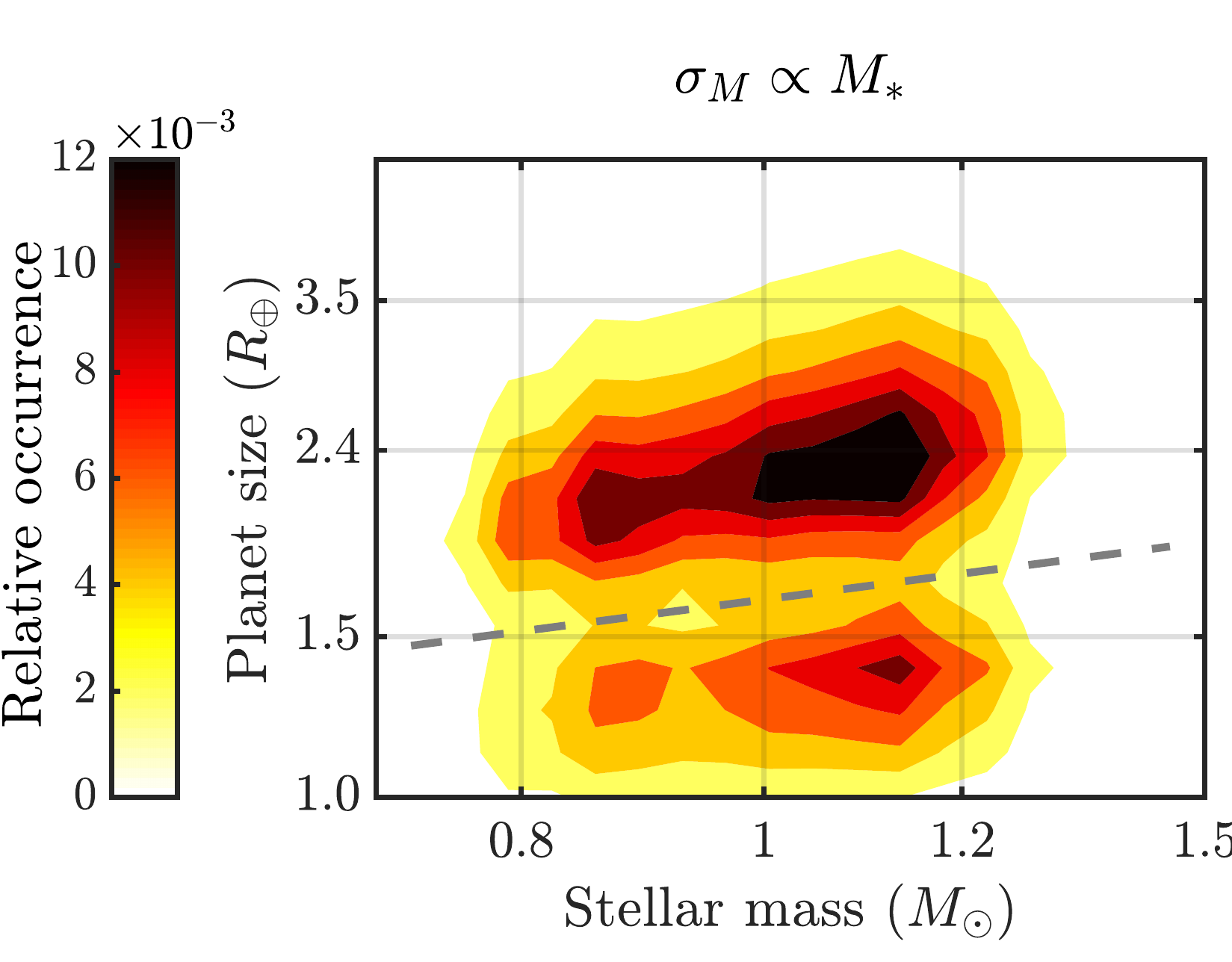}
\includegraphics[width=.32\textwidth,trim=0 0 0 0,clip]{Fulton_and_Petigura_2018__R_p_vs_M_s_600.png}
\caption{Comparison of core-powered mass-loss results (left and middle panel) with observations (right-panel) from \citet[][]{fulton2018a}. The left and middle panel show the resulting planet size distribution assuming no correlation between planet- and stellar-mass (left-panel) and for a linear correlation between the planet- and stellar-mass distributions (middle-panel), respectively. $\sigma_M$ corresponds to the peak in the planet mass distribution (see \Cref{eq:M_c_distr} for details). {The dashed lines indicate the slopes of the radius valley which are given by ${\text{d log}R_p}/{\text{d log} M_\ast}  \simeq 0.35$. Comparing the left and middle panels with the observations (right-panel) shows that assuming a linear-scaling between the planet- and stellar-mass distributions (middle-panel) and no-scaling (left-panel) both yield similar slopes which are both consistent with observations (left-panel). Core-powered mass-loss models only yield a slightly steeper slope when assuming a linear-scaling between the planet- and stellar-mass distributions compared to no-scaling because, to first order, the location of the radius valley is independent of the underlying planet mass distribution \citep[see Section 3.4 of][for details]{gupta2019a}.  As evident when comparing the left and middle panels, the intensities of the peaks above and below the valley are sensitive to the details of the planet mass distribution, but we are currently not able to favour one model over the other on this basis as the observations from \citet[][]{fulton2018a} that we are comparing to, in the right panel, have not been completeness-corrected.} As before, the ages, metallicities and masses of the host stars are modeled after the stellar distribution from the CKS dataset \citep{johnson2017a,fulton2017a}.}
\label{fig:stellar_prop_var_M_2}
\end{figure*}

\subsection{Investigating proposed correlations between the planet- and stellar-mass distributions}\label{sec:results_planet_stellar_mass}

As mentioned above, to reconcile the observed correlation between the location of the radius valley with stellar mass considering only atmospheric-loss due to photoevaporation, \citet{wu2019a} had to invoke a linear correlation between the planet- and stellar-mass distributions. Specifically, she inferred that the peak of the planet mass distribution, i.e., $\sigma_M$ (see \Cref{eq:M_c_distr}), scales linearly with the mass of the host star. 

In Sections \ref{sec:results_comparison_w_obs} and \ref{sec:results_stellar_prop_M}, we demonstrate that no such correlation is required when accounting for planetary evolution and atmospheric loss driven by the core-powered mass-loss mechanism. In other words, we find that in order to explain the observations we don't need to impose a  {linear} correlation between the planet- and stellar-mass distribution as inferred by \citet{wu2019a}  {for their photoevaporation model}. Nevertheless, in this section we investigate how our core-powered mass-loss results would change if we impose a linear relation between the planet- and stellar-mass distribution. Specifically, we assume that $\sigma_M/3 M_{\earth} = M_\ast/M_{\sun}$ (see \Cref{eq:M_c_distr} for details).

 {\Cref{fig:stellar_prop_var_M_2} demonstrates that the slope of the valley is similar when  assuming no correlation between the planet- and stellar-mass distributions (left-panel) or a linear correlation between the planet- and stellar-mass distributions (middle-panel). Furthermore, both models yield a good fit to observations from \citet[][]{fulton2018a} (right-panel) and both match a slope of ${\text{d log}R_p}/{\text{d log} M_\ast}  \simeq 0.35$ given by the dashed line in all three panels. Core-powered mass-loss models only yield a slightly steeper slope when assuming a linear-scaling between the planet- and stellar-mass distributions compared to no-scaling because, to first order, the location of the radius valley is independent of the underlying planet mass distribution (see Section 3.4 of \citet{gupta2019a} for details). As evident when comparing the left and middle panels, the intensities of the peaks above and below the valley are sensitive to the details of the planet mass distribution. However, we are currently not able to favour one model over the other on this basis because the observations from \citet[][]{fulton2018a} that we are comparing to in the right panel have not been completeness corrected for the relative intensities of the two peaks above and below the valley.}

 {In summary, since the slope of the radius valley is, to first order, not sensitive to the planet-mass distribution for the core-powered mass-loss mechanism, we do not find evidence for a linear correlation between the stellar- and planet-mass distributions, but we cannot rule it out either.}

\section{Discussion and Conclusions}\label{sec:conclusion}

In this paper we extended previous work on the core-powered mass-loss mechanism \citep{ginzburg2018a,gupta2019a}, and investigated how stellar mass, metallicity and age impact the resulting planet size distribution.

We first investigated the evolution of planets around a population of host stars modeled after the CKS dataset \citep{johnson2017a,fulton2017a}. This allows for a direct comparison between our core-powered mass-loss results and observations based on the CKS data set. We find that, to first order, the resulting planet size distribution as a function of period and/or stellar insolation is very similar to that obtained by \citet{gupta2019a}, who only  {considered} {`Sun-like'} host stars. However, modeling the actual host star properties yields, as expected, an even higher degree of agreement between core-powered mass-loss results and the observations \citep{fulton2018a}. In addition, we find that our core-powered mass-loss results are in excellent agreement with the observed planet size distribution as a function of both stellar mass \citep[e.g.,][]{fulton2018a,wu2019a} and metallicity \citep[e.g.,][]{owen2018a,petigura2018a,dong2018a,hirano2018a}. {Furthermore, our results also explain the shift in the planet size distribution to higher insolation flux as a function of stellar mass, as observed by \citet{fulton2018a}.}

In addition, we investigated how our core-powered mass-loss results depend on stellar mass, metallicity and age separately. This enabled us not only to understand the key physical processes responsible for the agreement of our results with observations, but also enables us to make predictions for the planet size distribution and its dependence on a wide rage of stellar parameters. 

We find that the planet size distribution varies with stellar mass because core-powered mass-loss depends on the bolometric luminosity which, in turn, is strongly correlated with stellar mass. Thus, more massive stars host planets with higher equilibrium temperature, making even more massive planets susceptible to complete atmospheric loss. We derived an analytical estimate for the slope of the valley in planet size and stellar mass parameter space and show that $\text{d log} R_p/ \text{d log} M_\ast \simeq (3\alpha-2)/36$, where $\alpha$ is the power-low index relating stellar mass and luminosity, such that $L_\ast/L_\odot = (M_\ast/M_\odot)^\alpha$. We find that $\alpha \simeq 4.6$ for the host stars in the CKS dataset and use this to calculate  that $\text{d log} R_p/ \text{d log} M_\ast \simeq 0.33$. This analytical estimate is in good agreement with our numerical results and the observations \citep{fulton2018a,wu2019a}. 

In contrast to stellar mass, we find that, to first order, that the core-powered mass-loss mechanism predicts a negligible dependence of the location of the radius valley as a function of metallicity and age. This is because, as shown in \citet{gupta2019a} and summarized in Section \ref{sec:results_stellar_prop_M}, the slope of the radius valley is, to first order, given by $ G M_p/c_s^2 R_{rcb} =$ constant, which has no explicit dependence on metallicity or age. Therefore, as long as the spontaneous mass-loss phase (boil-off phase) results in initial planet sizes $R_{rcb}= R_p \simeq  2R_c$ independent of envelope metallicity or age, we have that the slope of the valley, especially the lower edge that marks the upper contour of the super-Earth population, does not depend on metallicity or age. In reality, there is a weak dependence on age and metallicity resulting in slightly larger cores that are stripped as a function of time (see \Cref{fig:stellar_prop_var_A}) and slightly smaller cores that are stripped of their atmospheres as a function of metallicity (see \Cref{fig:stellar_prop_var_Z}). This dependence is due to the metallicity's effect on the mass-loss timescale. However, since this metallicity dependence does not appear in the exponent of the mass-loss timescales, the metallicity and hence age dependence is only a second-order effect and hence weak.

Unlike the location and slope of the radius valley, the size distribution of the sub-Neptune population does display a visible dependence on both metallicity and age. This is because the rate at which the envelope cools and contracts is related to its opacity. The size of sub-Neptunes therefore depends on both the opacity of the envelope and the age of the system. Assuming that the envelope opacity is proportional to the stellar metallicity, the cooling timescale is directly proportional to the stellar metallicity. As a result, sub-Neptunes will be larger around higher metallicity and/or younger stars than metal poor and/or older stars. We derived an analytical estimate for the size of sub-Neptunes as a function of metallicity and age and showed that $\text{d log} R_p/ \text{d log} Z_\ast \simeq 0.1$ and $\text{d log} R_p/ \text{d log} \tau_\ast \simeq -0.1$, respectively. 
Both of these analytical estimates are in agreement with our numerical results. In addition, the former is in good agreement with the observations of sub-Neptunes around metal-rich FGK stars \citep{owen2018a,dong2018a,petigura2018a} and M-dwarfs \citep{hirano2018a}; no direct comparison with observations can yet be made for the latter.

Although we just discussed here the dependence of the core-powered mass-loss mechanism on stellar mass, metallicity and age in isolation, some of these trends can be more pronounced in the observations because, for example, in the CKS dataset more massive stars are on average younger and more metal-rich.

\subsection{Observational test: Distinguishing between core-powered mass-loss and photoevaporation signatures}

Although both photoevaporation and core-powered mass-loss can explain the observed radius valley in the exoplanet size distribution and both mechanisms are likely to operate in conjunction in many systems, their different timescales and different dependencies on stellar properties provide a set of observational predictions that allow for the determination of their relative importance. Specifically, we make the following observational predictions: 

\subsubsection{Correlations between planet- and stellar-mass}

 {According to our core-powered mass-loss results, we find no evidence for a linear correlation between the host star and planet mass distributions. However, we cannot rule out such a linear correlation either, because the location of the radius valley in the core-powered mass-loss scenario does not strongly depend on planet mass \citep[see Section 3.4 from][for details]{gupta2019a}.} 

\subsubsection{Slope of the radius valley as a function of stellar mass (or luminosity)}
Since the positive correlation between the location of the radius valley with stellar mass is mainly due to the stellar mass-luminosity relation, the core-powered mass-loss model can be tested by examining the slope of the radius valley for different stellar masses. For example, as shown above, the slope of the radius valley as a function of stellar mass is given by $\text{d log}R_p/\text{d log}M_\ast = (3 \alpha - 2)/36 $, where $L_\ast/L_{\sun}=(M_\ast/M_{\sun})^{\alpha}$. The slope will therefore be steeper for planets around host stars with larger $\alpha$ and vice-versa. Best of all, one can examine planets around two host star populations with different $\alpha$ ($\alpha$ can be directly determined from the host stars in the sample) and check for the predicted change in slope of the radius valley as a function of stellar mass (or luminosity).

\subsubsection{Relative abundance of super-Earths and sub-Neptunes as a function of age}
Since atmospheric mass-loss driven by the core-powered mass-loss mechanism proceeds over Gyr timescales, we predict that the relative abundance of super-Earths with respect to sub-Neptunes increases with age. Specifically, we expect the abundance of super-Earths to continue to increases significantly after 500 Myrs, since typical atmospheric mass-loss timescales are of the order of a Gyr. In contrast, since photoevaporation is driven by the X-ray and EUV radiation from the host star, which declines drastically after the first 100 Myrs, the relative abundance of super-Earths should remain constant after the first 100 Myrs. Photoevaporation models therefore predict no significant increase in the super-Earth population after the first 100-200 Myrs.

\subsubsection{Planets in the gap}
Since planets continue to lose mass over Gyr timescales, the core-powered mass-loss mechanism predicts that some planets will be still be losing mass today. Observationally these planets can be caught as they cross the radius valley, i.e. they maybe found inside the gap. In addition, atmospheric mass-loss may be directly detectable observationally.

\subsection{Outlook}
In this work we have shown how planetary evolution under the core-powered mass-loss mechanism can explain many of the observed features of the distribution of small, short-period exoplanets and their dependence on stellar properties. Although we have just summarized a list of observational test that should be able to distinguish between signatures imprinted on the exoplanet population by photoevaporation and core-powered mass-loss, we would like to emphasize that it appears almost inevitable that both processes take place to some extent. We therefore suspect that they both play a role in shaping the observed super-Earth and sub-Neptune populations. To that end, we plan to combine photoevaporation and core-powered mass-loss models in future work. Combining both of these mass-loss mechanisms may turn out to be crucial for correctly inferring physical properties (e.g., core densities) of the exoplanet population from the observations.

\section*{Acknowledgements}
We are grateful to the California-\textit{Kepler} Survey team for making their data publicly available. {
We thank James Owen for valuable comments that helped to improve the manuscript.} {A.G. also thanks Erik Petigura and B.J. Fulton for useful discussions related to this work.} H.E.S. gratefully acknowledges support from the National Aeronautics and Space Administration under grant No. $17~\rm{XRP}17\_~2-0055$ issued through the Exoplanet Research Program.

\bibliographystyle{mnras}
\bibliography{planet_evo}

\begin{thebibliography}{}
\makeatletter
\relax
\def\mn@urlcharsother{\let\do\@makeother \do\$\do\&\do\#\do\^\do\_\do\%\do\~}
\def\mn@doi{\begingroup\mn@urlcharsother \@ifnextchar [ {\mn@doi@}
  {\mn@doi@[]}}
\def\mn@doi@[#1]#2{\def\@tempa{#1}\ifx\@tempa\@empty \href
  {http://dx.doi.org/#2} {doi:#2}\else \href {http://dx.doi.org/#2} {#1}\fi
  \endgroup}
\def\mn@eprint#1#2{\mn@eprint@#1:#2::\@nil}
\def\mn@eprint@arXiv#1{\href {http://arxiv.org/abs/#1} {{\tt arXiv:#1}}}
\def\mn@eprint@dblp#1{\href {http://dblp.uni-trier.de/rec/bibtex/#1.xml}
  {dblp:#1}}
\def\mn@eprint@#1:#2:#3:#4\@nil{\def\@tempa {#1}\def\@tempb {#2}\def\@tempc
  {#3}\ifx \@tempc \@empty \let \@tempc \@tempb \let \@tempb \@tempa \fi \ifx
  \@tempb \@empty \def\@tempb {arXiv}\fi \@ifundefined
  {mn@eprint@\@tempb}{\@tempb:\@tempc}{\expandafter \expandafter \csname
  mn@eprint@\@tempb\endcsname \expandafter{\@tempc}}}

\bibitem[\protect\citeauthoryear{{Berger}, {Huber}, {Gaidos}  \& {van
  Saders}}{{Berger} et~al.}{2018}]{berger2018a}
{Berger} T.~A.,  {Huber} D.,  {Gaidos} E.,   {van Saders} J.~L.,  2018, \mn@doi
  [\apj] {10.3847/1538-4357/aada83}, \href
  {https://ui.adsabs.harvard.edu/\#abs/2018ApJ...866...99B} {866, 99}

\bibitem[\protect\citeauthoryear{{Borucki} et~al.,}{{Borucki}
  et~al.}{2010}]{BK10}
{Borucki} W.~J.,  et~al., 2010, \mn@doi [Science] {10.1126/science.1185402},
  \href {http://adsabs.harvard.edu/abs/2010Sci...327..977B} {327, 977}

\bibitem[\protect\citeauthoryear{{Bower}, {Kitzmann}, {Wolf}, {Sanan}, {Dorn}
  \& {Oza}}{{Bower} et~al.}{2019}]{bower2019}
{Bower} D.~J.,  {Kitzmann} D.,  {Wolf} A.~S.,  {Sanan} P.,  {Dorn} C.,   {Oza}
  A.~V.,  2019, arXiv e-prints, \href
  {https://ui.adsabs.harvard.edu/abs/2019arXiv190408300B} {p. arXiv:1904.08300}

\bibitem[\protect\citeauthoryear{{Brown}, {Latham}, {Everett}  \&
  {Esquerdo}}{{Brown} et~al.}{2011}]{brown2011a}
{Brown} T.~M.,  {Latham} D.~W.,  {Everett} M.~E.,   {Esquerdo} G.~A.,  2011,
  \mn@doi [\aj] {10.1088/0004-6256/142/4/112}, \href
  {https://ui.adsabs.harvard.edu/\#abs/2011AJ....142..112B} {142, 112}

\bibitem[\protect\citeauthoryear{{Carter} et~al.,}{{Carter}
  et~al.}{2012}]{carter2012a}
{Carter} J.~A.,  et~al., 2012, \mn@doi [Science] {10.1126/science.1223269},
  \href {https://ui.adsabs.harvard.edu/\#abs/2012Sci...337..556C} {337, 556}

\bibitem[\protect\citeauthoryear{{Chen} \& {Rogers}}{{Chen} \&
  {Rogers}}{2016}]{chen2016a}
{Chen} H.,  {Rogers} L.~A.,  2016, \mn@doi [\apj]
  {10.3847/0004-637X/831/2/180}, \href
  {https://ui.adsabs.harvard.edu/\#abs/2016ApJ...831..180C} {831, 180}

\bibitem[\protect\citeauthoryear{{Christiansen} et~al.,}{{Christiansen}
  et~al.}{2015}]{christiansen2015a}
{Christiansen} J.~L.,  et~al., 2015, \mn@doi [\apj]
  {10.1088/0004-637X/810/2/95}, \href
  {https://ui.adsabs.harvard.edu/abs/2015ApJ...810...95C} {810, 95}

\bibitem[\protect\citeauthoryear{{Dong}, {Xie}, {Zhou}, {Zheng}  \&
  {Luo}}{{Dong} et~al.}{2018}]{dong2018a}
{Dong} S.,  {Xie} J.-W.,  {Zhou} J.-L.,  {Zheng} Z.,   {Luo} A.,  2018, \mn@doi
  [Proceedings of the National Academy of Science] {10.1073/pnas.1711406115},
  \href {https://ui.adsabs.harvard.edu/\#abs/2018PNAS..115..266D} {115, 266}

\bibitem[\protect\citeauthoryear{{Dorn}, {Harrison}, {Bonsor}  \&
  {Hands}}{{Dorn} et~al.}{2019}]{dorn2019a}
{Dorn} C.,  {Harrison} J.~H.~D.,  {Bonsor} A.,   {Hands} T.~O.,  2019, \mn@doi
  [\mnras] {10.1093/mnras/sty3435}, \href
  {https://ui.adsabs.harvard.edu/abs/2019MNRAS.484..712D} {484, 712}

\bibitem[\protect\citeauthoryear{{Dressing} et~al.,}{{Dressing}
  et~al.}{2015}]{dressing2015a}
{Dressing} C.~D.,  et~al., 2015, \mn@doi [\apj] {10.1088/0004-637X/800/2/135},
  \href {https://ui.adsabs.harvard.edu/abs/2015ApJ...800..135D} {800, 135}

\bibitem[\protect\citeauthoryear{{Eker} et~al.,}{{Eker}
  et~al.}{2018}]{eker2018a}
{Eker} Z.,  et~al., 2018, \mn@doi [\mnras] {10.1093/mnras/sty1834}, \href
  {http://adsabs.harvard.edu/abs/2018MNRAS.479.5491E} {479, 5491}

\bibitem[\protect\citeauthoryear{{Fortney}, {Marley}  \& {Barnes}}{{Fortney}
  et~al.}{2007}]{fortney2007a}
{Fortney} J.~J.,  {Marley} M.~S.,   {Barnes} J.~W.,  2007, \mn@doi [\apj]
  {10.1086/512120}, \href {http://adsabs.harvard.edu/abs/2007ApJ...659.1661F}
  {659, 1661}

\bibitem[\protect\citeauthoryear{{Freedman}, {Marley}  \& {Lodders}}{{Freedman}
  et~al.}{2008}]{freedman2008a}
{Freedman} R.~S.,  {Marley} M.~S.,   {Lodders} K.,  2008, \mn@doi [\apjs]
  {10.1086/521793}, \href {http://adsabs.harvard.edu/abs/2008ApJS..174..504F}
  {174, 504}

\bibitem[\protect\citeauthoryear{{Fressin} et~al.,}{{Fressin}
  et~al.}{2013}]{FT13}
{Fressin} F.,  et~al., 2013, \mn@doi [\apj] {10.1088/0004-637X/766/2/81}, \href
  {http://adsabs.harvard.edu/abs/2013ApJ...766...81F} {766, 81}

\bibitem[\protect\citeauthoryear{{Fulton} \& {Petigura}}{{Fulton} \&
  {Petigura}}{2018}]{fulton2018a}
{Fulton} B.~J.,  {Petigura} E.~A.,  2018, \mn@doi [\aj]
  {10.3847/1538-3881/aae828}, \href
  {https://ui.adsabs.harvard.edu/\#abs/2018AJ....156..264F} {156, 264}

\bibitem[\protect\citeauthoryear{{Fulton} et~al.,}{{Fulton}
  et~al.}{2017}]{fulton2017a}
{Fulton} B.~J.,  et~al., 2017, \mn@doi [\aj] {10.3847/1538-3881/aa80eb}, \href
  {http://adsabs.harvard.edu/abs/2017AJ....154..109F} {154, 109}

\bibitem[\protect\citeauthoryear{{Ginzburg}, {Schlichting}  \&
  {Sari}}{{Ginzburg} et~al.}{2016}]{ginzburg2016a}
{Ginzburg} S.,  {Schlichting} H.~E.,   {Sari} R.,  2016, \mn@doi [\apj]
  {10.3847/0004-637X/825/1/29}, \href
  {http://adsabs.harvard.edu/abs/2016ApJ...825...29G} {825, 29}

\bibitem[\protect\citeauthoryear{{Ginzburg}, {Schlichting}  \&
  {Sari}}{{Ginzburg} et~al.}{2018}]{ginzburg2018a}
{Ginzburg} S.,  {Schlichting} H.~E.,   {Sari} R.,  2018, \mn@doi [\mnras]
  {10.1093/mnras/sty290}, \href
  {http://adsabs.harvard.edu/abs/2018MNRAS.476..759G} {476, 759}

\bibitem[\protect\citeauthoryear{{Gupta} \& {Schlichting}}{{Gupta} \&
  {Schlichting}}{2019}]{gupta2019a}
{Gupta} A.,  {Schlichting} H.~E.,  2019, \mn@doi [\mnras]
  {10.1093/mnras/stz1230}, \href
  {https://ui.adsabs.harvard.edu/abs/2019MNRAS.487...24G} {487, 24}

\bibitem[\protect\citeauthoryear{{Hadden} \& {Lithwick}}{{Hadden} \&
  {Lithwick}}{2017}]{hadden2017a}
{Hadden} S.,  {Lithwick} Y.,  2017, \mn@doi [\aj] {10.3847/1538-3881/aa71ef},
  \href {https://ui.adsabs.harvard.edu/\#abs/2017AJ....154....5H} {154, 5}

\bibitem[\protect\citeauthoryear{{Hirano} et~al.,}{{Hirano}
  et~al.}{2018}]{hirano2018a}
{Hirano} T.,  et~al., 2018, \mn@doi [\aj] {10.3847/1538-3881/aaa9c1}, \href
  {https://ui.adsabs.harvard.edu/\#abs/2018AJ....155..127H} {155, 127}

\bibitem[\protect\citeauthoryear{{Inamdar} \& {Schlichting}}{{Inamdar} \&
  {Schlichting}}{2015}]{IS15}
{Inamdar} N.~K.,  {Schlichting} H.~E.,  2015, \mn@doi [\mnras]
  {10.1093/mnras/stv030}, \href
  {http://adsabs.harvard.edu/abs/2015MNRAS.448.1751I} {448, 1751}

\bibitem[\protect\citeauthoryear{{Jackson}, {Davis}  \& {Wheatley}}{{Jackson}
  et~al.}{2012}]{jackson2012a}
{Jackson} A.~P.,  {Davis} T.~A.,   {Wheatley} P.~J.,  2012, \mn@doi [\mnras]
  {10.1111/j.1365-2966.2012.20657.x}, \href
  {https://ui.adsabs.harvard.edu/\#abs/2012MNRAS.422.2024J} {422, 2024}

\bibitem[\protect\citeauthoryear{{Jin} \& {Mordasini}}{{Jin} \&
  {Mordasini}}{2018}]{jin2018a}
{Jin} S.,  {Mordasini} C.,  2018, \mn@doi [\apj] {10.3847/1538-4357/aa9f1e},
  \href {https://ui.adsabs.harvard.edu/\#abs/2018ApJ...853..163J} {853, 163}

\bibitem[\protect\citeauthoryear{{Jin}, {Mordasini}, {Parmentier}, {van
  Boekel}, {Henning}  \& {Ji}}{{Jin} et~al.}{2014}]{jin2014a}
{Jin} S.,  {Mordasini} C.,  {Parmentier} V.,  {van Boekel} R.,  {Henning} T.,
  {Ji} J.,  2014, \mn@doi [\apj] {10.1088/0004-637X/795/1/65}, \href
  {https://ui.adsabs.harvard.edu/\#abs/2014ApJ...795...65J} {795, 65}

\bibitem[\protect\citeauthoryear{{Johnson} et~al.,}{{Johnson}
  et~al.}{2017}]{johnson2017a}
{Johnson} J.~A.,  et~al., 2017, \mn@doi [\aj] {10.3847/1538-3881/aa80e7}, \href
  {http://adsabs.harvard.edu/abs/2017AJ....154..108J} {154, 108}

\bibitem[\protect\citeauthoryear{{Lee} \& {Chiang}}{{Lee} \&
  {Chiang}}{2015}]{lee2015a}
{Lee} E.~J.,  {Chiang} E.,  2015, \mn@doi [\apj] {10.1088/0004-637X/811/1/41},
  \href {https://ui.adsabs.harvard.edu/\#abs/2015ApJ...811...41L} {811, 41}

\bibitem[\protect\citeauthoryear{{Lopez} \& {Fortney}}{{Lopez} \&
  {Fortney}}{2013}]{LF13}
{Lopez} E.~D.,  {Fortney} J.~J.,  2013, \mn@doi [\apj]
  {10.1088/0004-637X/776/1/2}, \href
  {http://adsabs.harvard.edu/abs/2013ApJ...776....2L} {776, 2}

\bibitem[\protect\citeauthoryear{{Lopez} \& {Fortney}}{{Lopez} \&
  {Fortney}}{2014}]{LF14}
{Lopez} E.~D.,  {Fortney} J.~J.,  2014, \mn@doi [\apj]
  {10.1088/0004-637X/792/1/1}, \href
  {http://adsabs.harvard.edu/abs/2014ApJ...792....1L} {792, 1}

\bibitem[\protect\citeauthoryear{{Mann} et~al.,}{{Mann}
  et~al.}{2016}]{mann2016a}
{Mann} A.~W.,  et~al., 2016, \mn@doi [\apj] {10.3847/0004-637X/818/1/46}, \href
  {https://ui.adsabs.harvard.edu/abs/2016ApJ...818...46M} {818, 46}

\bibitem[\protect\citeauthoryear{{Marcy}, {Weiss}, {Petigura}, {Isaacson},
  {Howard}  \& {Buchhave}}{{Marcy} et~al.}{2014a}]{marcy2014b}
{Marcy} G.~W.,  {Weiss} L.~M.,  {Petigura} E.~A.,  {Isaacson} H.,  {Howard}
  A.~W.,   {Buchhave} L.~A.,  2014a, \mn@doi [Proceedings of the National
  Academy of Science] {10.1073/pnas.1304197111}, \href
  {https://ui.adsabs.harvard.edu/\#abs/2014PNAS..11112655M} {111, 12655}

\bibitem[\protect\citeauthoryear{{Marcy} et~al.,}{{Marcy}
  et~al.}{2014b}]{marcy2014a}
{Marcy} G.~W.,  et~al., 2014b, \mn@doi [\apjs] {10.1088/0067-0049/210/2/20},
  \href {http://adsabs.harvard.edu/abs/2014ApJS..210...20M} {210, 20}

\bibitem[\protect\citeauthoryear{{Martinez}, {Cunha}, {Ghezzi}  \&
  {Smith}}{{Martinez} et~al.}{2019}]{martinez2019a}
{Martinez} C.~F.,  {Cunha} K.,  {Ghezzi} L.,   {Smith} V.~V.,  2019, \mn@doi
  [\apj] {10.3847/1538-4357/ab0d93}, \href
  {http://adsabs.harvard.edu/abs/2019ApJ...875...29M} {875, 29}

\bibitem[\protect\citeauthoryear{{Owen} \& {Murray-Clay}}{{Owen} \&
  {Murray-Clay}}{2018}]{owen2018a}
{Owen} J.~E.,  {Murray-Clay} R.,  2018, \mn@doi [\mnras]
  {10.1093/mnras/sty1943}, \href
  {http://adsabs.harvard.edu/abs/2018MNRAS.480.2206O} {480, 2206}

\bibitem[\protect\citeauthoryear{{Owen} \& {Wu}}{{Owen} \&
  {Wu}}{2013}]{owen2013a}
{Owen} J.~E.,  {Wu} Y.,  2013, \mn@doi [\apj] {10.1088/0004-637X/775/2/105},
  \href {http://adsabs.harvard.edu/abs/2013ApJ...775..105O} {775, 105}

\bibitem[\protect\citeauthoryear{{Owen} \& {Wu}}{{Owen} \&
  {Wu}}{2016}]{owen2016a}
{Owen} J.~E.,  {Wu} Y.,  2016, \mn@doi [\apj] {10.3847/0004-637X/817/2/107},
  \href {http://adsabs.harvard.edu/abs/2016ApJ...817..107O} {817, 107}

\bibitem[\protect\citeauthoryear{{Owen} \& {Wu}}{{Owen} \&
  {Wu}}{2017}]{owen2017a}
{Owen} J.~E.,  {Wu} Y.,  2017, \mn@doi [\apj] {10.3847/1538-4357/aa890a}, \href
  {http://adsabs.harvard.edu/abs/2017ApJ...847...29O} {847, 29}

\bibitem[\protect\citeauthoryear{{Petigura}, {Marcy}  \& {Howard}}{{Petigura}
  et~al.}{2013}]{PM13}
{Petigura} E.~A.,  {Marcy} G.~W.,   {Howard} A.~W.,  2013, \mn@doi [\apj]
  {10.1088/0004-637X/770/1/69}, \href
  {http://adsabs.harvard.edu/abs/2013ApJ...770...69P} {770, 69}

\bibitem[\protect\citeauthoryear{{Petigura} et~al.,}{{Petigura}
  et~al.}{2017}]{petigura2017a}
{Petigura} E.~A.,  et~al., 2017, \mn@doi [\aj] {10.3847/1538-3881/aa80de},
  \href {http://adsabs.harvard.edu/abs/2017AJ....154..107P} {154, 107}

\bibitem[\protect\citeauthoryear{{Petigura} et~al.,}{{Petigura}
  et~al.}{2018}]{petigura2018a}
{Petigura} E.~A.,  et~al., 2018, \mn@doi [\aj] {10.3847/1538-3881/aaa54c},
  \href {https://ui.adsabs.harvard.edu/\#abs/2018AJ....155...89P} {155, 89}

\bibitem[\protect\citeauthoryear{{Piso} \& {Youdin}}{{Piso} \&
  {Youdin}}{2014}]{PY14}
{Piso} A.-M.~A.,  {Youdin} A.~N.,  2014, \mn@doi [\apj]
  {10.1088/0004-637X/786/1/21}, \href
  {http://adsabs.harvard.edu/abs/2014ApJ...786...21P} {786, 21}

\bibitem[\protect\citeauthoryear{{Rafikov}}{{Rafikov}}{2006}]{rafikov2006a}
{Rafikov} R.~R.,  2006, \mn@doi [\apj] {10.1086/505695}, \href
  {https://ui.adsabs.harvard.edu/abs/2006ApJ...648..666R} {648, 666}

\bibitem[\protect\citeauthoryear{{Rizzuto}, {Vanderburg}, {Mann}, {Kraus},
  {Dressing}, {Ag{\"u}eros}, {Douglas}  \& {Krolikowski}}{{Rizzuto}
  et~al.}{2018}]{rizzuto2018a}
{Rizzuto} A.~C.,  {Vanderburg} A.,  {Mann} A.~W.,  {Kraus} A.~L.,  {Dressing}
  C.~D.,  {Ag{\"u}eros} M.~A.,  {Douglas} S.~T.,   {Krolikowski} D.~M.,  2018,
  \mn@doi [\aj] {10.3847/1538-3881/aadf37}, \href
  {https://ui.adsabs.harvard.edu/abs/2018AJ....156..195R} {156, 195}

\bibitem[\protect\citeauthoryear{{Rogers}}{{Rogers}}{2015}]{rogers2015a}
{Rogers} L.~A.,  2015, \mn@doi [\apj] {10.1088/0004-637X/801/1/41}, \href
  {http://adsabs.harvard.edu/abs/2015ApJ...801...41R} {801, 41}

\bibitem[\protect\citeauthoryear{{Tu}, {Johnstone}, {G{\"u}del}  \&
  {Lammer}}{{Tu} et~al.}{2015}]{tu2015a}
{Tu} L.,  {Johnstone} C.~P.,  {G{\"u}del} M.,   {Lammer} H.,  2015, \mn@doi
  [\aap] {10.1051/0004-6361/201526146}, \href
  {https://ui.adsabs.harvard.edu/\#abs/2015A&A...577L...3T} {577, L3}

\bibitem[\protect\citeauthoryear{{Valencia}, {O'Connell}  \&
  {Sasselov}}{{Valencia} et~al.}{2006}]{valencia2006a}
{Valencia} D.,  {O'Connell} R.~J.,   {Sasselov} D.,  2006, \mn@doi [\icarus]
  {10.1016/j.icarus.2005.11.021}, \href
  {http://adsabs.harvard.edu/abs/2006Icar..181..545V} {181, 545}

\bibitem[\protect\citeauthoryear{{Van Eylen}, {Agentoft}, {Lundkvist},
  {Kjeldsen}, {Owen}, {Fulton}, {Petigura}  \& {Snellen}}{{Van Eylen}
  et~al.}{2018}]{eylen2018a}
{Van Eylen} V.,  {Agentoft} C.,  {Lundkvist} M.~S.,  {Kjeldsen} H.,  {Owen}
  J.~E.,  {Fulton} B.~J.,  {Petigura} E.,   {Snellen} I.,  2018, \mn@doi
  [\mnras] {10.1093/mnras/sty1783}, \href
  {http://adsabs.harvard.edu/abs/2018MNRAS.479.4786V} {479, 4786}

\bibitem[\protect\citeauthoryear{{Wallack} et~al.,}{{Wallack}
  et~al.}{2019}]{wallack2019a}
{Wallack} N.~L.,  et~al., 2019, \mn@doi [\aj] {10.3847/1538-3881/ab2a05}, \href
  {https://ui.adsabs.harvard.edu/abs/2019AJ....158..217W} {158, 217}

\bibitem[\protect\citeauthoryear{{Weiss} \& {Marcy}}{{Weiss} \&
  {Marcy}}{2014}]{weiss2014a}
{Weiss} L.~M.,  {Marcy} G.~W.,  2014, \mn@doi [\apj]
  {10.1088/2041-8205/783/1/L6}, \href
  {https://ui.adsabs.harvard.edu/\#abs/2014ApJ...783L...6W} {783, L6}

\bibitem[\protect\citeauthoryear{{Wu}}{{Wu}}{2019}]{wu2019a}
{Wu} Y.,  2019, \mn@doi [\apj] {10.3847/1538-4357/ab06f8}, \href
  {https://ui.adsabs.harvard.edu/abs/2019ApJ...874...91W} {874, 91}

\makeatother
\end{thebibliography}

\bsp	
\label{lastpage}
\end{document}